\documentclass[superscriptaddress,prx,aps,longbibliography,twocolumn]{revtex4-2}
\usepackage{tikz}  
\usetikzlibrary{shapes,arrows,positioning,automata,backgrounds,calc,er,patterns}

\usepackage{graphicx}
\usepackage{dcolumn}
\usepackage{bm}
\usepackage{amsmath}
\usepackage{amssymb}
\usepackage{amsthm}
\usepackage{amsfonts}
\usepackage{enumerate}
\usepackage{array}
\usepackage{changes}
\newcolumntype{P}[1]{>{\centering\arraybackslash}p{#1}}
\usepackage{float}
\usepackage{dsfont}

\newcommand{\ket}[1]{\left|#1\right\rangle}
\newcommand{\bra}[1]{\left\langle#1\right|}
\newcommand{\braket}[2]{\left\langle#1\mid#2\right\rangle}
\newcommand{\sx}{\hat{\sigma}^x}
\newcommand{\sy}{\hat{\sigma}^y}
\newcommand{\sz}{\hat{\sigma}^z}

\usepackage[colorlinks=true,urlcolor=blue,citecolor=blue,allcolors=blue]{hyperref}
\usepackage{cleveref}

\begin{document}
\title{Dynamical learning and quantum memory with  non-Hermitian many-body systems}
%\title{Imaginary quantum systems with real memory}

\newcommand{\Aalto}{MSP Group, QTF Centre of Excellence, Department of Applied Physics,\\
Aalto University, FI-00076 Aalto, Espoo, Finland}
\newcommand{\ICMM}{Interdisciplinary Centre for Mathematical Modelling and Department of Mathematical Sciences,\\ Loughborough University, Loughborough, Leicestershire LE11 3TU, United Kingdom}
\newcommand{\LUPhys}{Department of Physics, Loughborough University, Loughborough, LE11 3TU, United Kingdom}
% Authors
\author{Moein N. Ivaki}
\email{moein.najafiivaki@aalto.fi}
\address{\Aalto}
\author{Austin J. Szuminsky}
\address{\Aalto}
\author{Achilleas Lazarides}
\address{\ICMM}
\author{Alexandre Zagoskin}
\address{\LUPhys}
\author{Gerard McCaul}
\address{\LUPhys}
\author{Tapio Ala-Nissila}
\email[Corresponding author: \vspace{-3pt}]{tapio.ala-nissila@aalto.fi}

\address{\Aalto}
\address{\ICMM}
\date{\today}

\begin{abstract}

Non-Hermitian (NH) systems provide a fertile platform for quantum technologies, owing in part to their distinct dynamical phases. These systems can be characterized by the preservation or spontaneous breaking of parity-time reversal symmetry, significantly impacting the dynamical behavior of quantum resources such as entanglement and purity; resources which in turn govern the system's information processing and memory capacity. Here we investigate this relationship using the example of an interacting NH spin system defined on random graphs. We show that the onset of the first exceptional point - marking the real-to-complex spectral transition - also corresponds to an abrupt change in the system’s learning capacity. We further demonstrate that this transition is controllable via local disorder and spin interactions strength, thereby defining a tunable learnability threshold. Within the learning phase, the system exhibits the key features required for memory-dependent reservoir computing. This makes explicit a direct link between spectral structure and computational capacity, further establishing non-Hermiticity, and more broadly engineered dissipation, as a dynamic resource for temporal quantum machine learning.

\end{abstract}
\maketitle
% #######################################################################
% #######################################################################
\section{Introduction}
The key quantity in all learning processes is memory. This is true of both natural and synthetic agents: The capacity to generalize from experience relies on the ability to store it. Given however that any computation associated with learning must take place on a physical substrate~\cite{Lloyd2002}, it is natural to ask what dynamical processes enable it. Whether realized in neural tissue, silicon circuits, or quantum media, computation is ultimately an interpretive structure superimposed on the dynamics that realize it. From this perspective, the pursuit of practical learning architectures becomes a search for dynamical systems capable of encoding, transforming, and recalling information with precision and efficiency. In this hunt, quantum systems present one of the richest targets. The exponential scaling of their underlying state space, long-range nonclassical correlations, and dependence on initial conditions give them the potential to process high-dimensional data far more efficiently than classical methods. It is for this reason that Quantum Machine Learning (QML) seeks to harness these unique features, applied to learning tasks such as optimization, classification, and prediction~\cite{Biamonte2017,Schuld2015,dunjko2018machine,liu2021rigorous}.

Among various QML paradigms, quantum reservoir computing (QRC)~\cite{PhysRevApplied.8.024030,Fujii2020,Govia2021,PhysRevX.11.041062,kornjavca2024large,Ghosh2019QRC,mujal2021opportunities,hu2024overcoming,senanian2024microwave,martinez2023information} is particularly well-suited to near-term devices. Inspired by classical reservoir computing and recurrent neural networks~\cite{Tanaka2018,Ozturk2007,lukovsevivcius2009reservoir,butcher2013reservoir,vrugt2024introduction,PhysRevX.7.011015,lim2020predicting}, QRC offloads the burden of processing information onto the untrained dynamics of a physical system, dubbed the \textit{reservoir}. The rationale for doing so lies in the fact that \textit{any} input-output relationship can be linearized, provided the input data is embedded into a sufficiently high dimensional feature space~\cite{kutz_dynamic_2016}. By encoding inputs into the dynamics of quantum systems, it is only necessary to train a single linear computational layer, which translates measurements read out from the reservoir into computational outputs. QRC therefore minimizes the need for extensive network training, making it a highly efficient approach compared to conventional variational algorithms~\cite{cerezo2021variational,larocca2025barren}. Such quantum learning scheme is inherently suited to analog quantum devices, as it is agnostic to the precise form of the reservoir dynamics. As such, it bypasses the stringent control requirements of gate-based circuits, offering both hardware efficiency and robustness to noise.  Hence, this family of learning approaches aligns well near-terms goals of implementation in noisy intermediate-scale quantum devices~\cite{RevModPhys.94.015004}. QRC has been applied to a range of problems, including quantum state classification, tomography, chaotic system modeling and non-linear processing~\cite{PhysRevLett.127.260401, palacios2024role,mujal2023time,settino2024memory,PRXQuantum.5.040325,sornsaeng2024quantum}.

Broadly speaking, there are three properties that enable a reservoir to be used for universal computations. These are fading memory, input separability, and the echo-state property~\cite{chen2019learning,sannia2024dissipation,Grigoryeva2018,nokkala2021gaussian}. There is however an inherent difficulty in satisfying these desiderata in the context of quantum dynamics. In particular, fading memory becomes an impossibility under the restriction of unitary operations and global projective measurement~\cite{McCaul2025Minimal}. This is an essential challenge in realizing QRC, as in many tasks successful learning requires memory. This dependence is unsurprising however. To paraphrase a famous Albert: learning is what remains when all else is forgotten~\cite{Einstein2005Ideas}. In this context, dynamical measures of chaoticity and lossiness can be tied directly to the memory capacity and precision of a given temporal learning task~\cite{xia2022reservoir, ivaki2024quantum,Martinez-Pena2021}. 

For this reason, the characterization and control of dynamical phases of quantum systems is vital to realizing the mathematical features quantum learning algorithms rely on ~\cite{domingo2023taking,fry2023optimizing,harrington2022engineered,PhysRevA.110.042416,olivera2023benefits,PhysRevA.111.012622}. A natural approach to introducing the necessary dynamical map is via \textit{partial} trace operations, in which one repeatedly writes and erases an input locally in an otherwise unitary system. However, such trace-based mechanisms, as originally proposed in early formulations of QRC~\cite{PhysRevApplied.8.024030}, offer little flexibility in parameterizing and tuning. This is problematic, as optimizing for generalized learning tasks can only be achieved with parametrized control over the dynamical processes generating it. This gives rise to two key objectives: designing simple, parameterizable dynamical models with tunable memory effects, and developing reliable schemes for their physical implementations.

\begin{figure}[t!]
\centering
\includegraphics[width=0.99\linewidth]{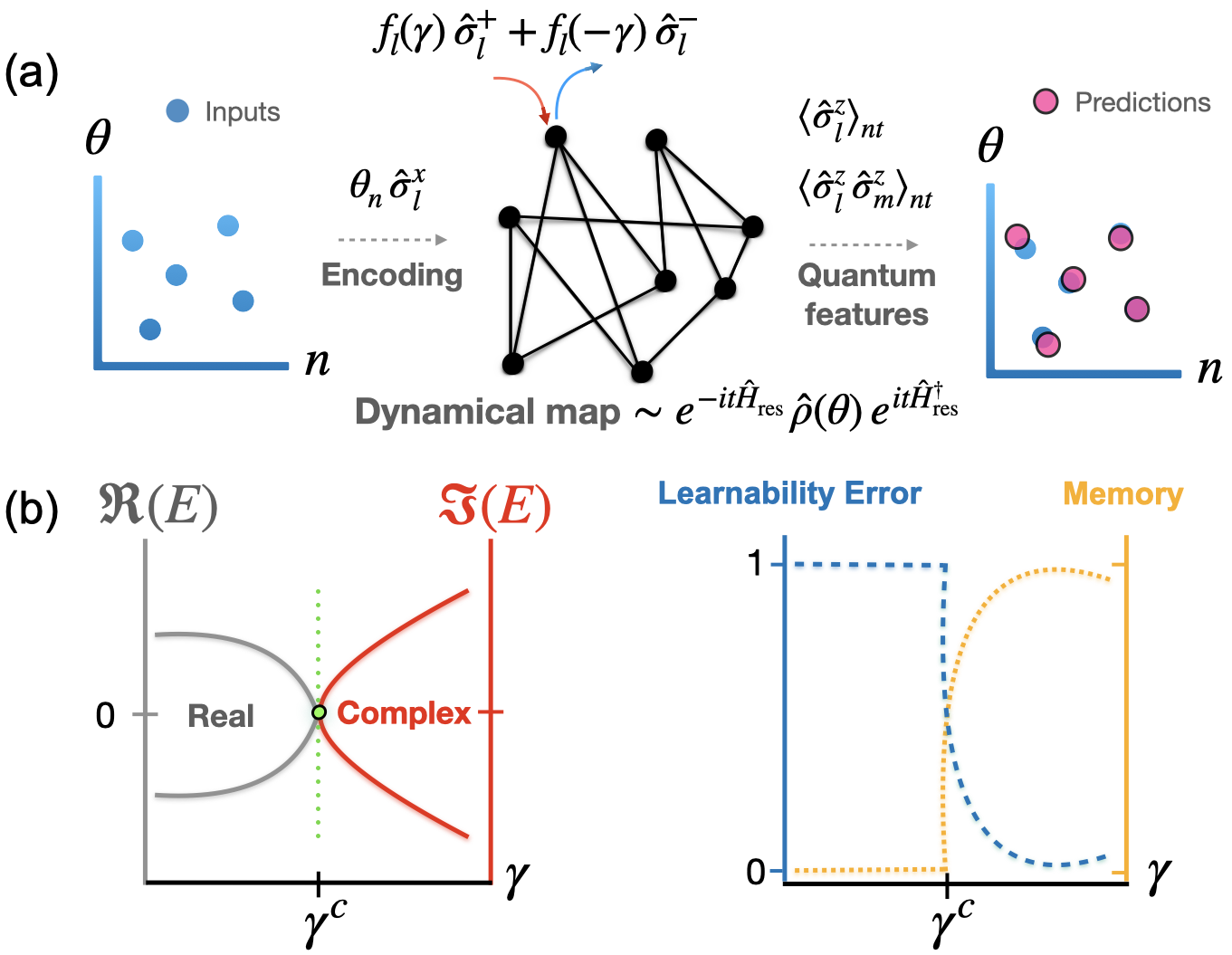}
\caption{ \textbf{Dynamical learning in a lossy quantum system}. \textbf{(a)} An input sequence ${\theta_n}$ is locally encoded through $x$-axis rotations at each step $n$, and applied to all nodes of a non-Hermitian quantum spin system defined on a random graph. External perturbations manifest as asymmetric spin flip operations at each site $l$, modeled by $f_l(\gamma)\,\hat{\sigma}_l^+$ and $f_l(-\gamma)\,\hat{\sigma}_l^-$. In our setup, this function is taken to be linear in the perturbation strength $f_l(\pm\gamma) \sim h_l \pm \gamma$. The current input-dependent state of the system is represented by a density matrix $\hat{\rho}(\theta)$. After nonunitary evolution under the Hamiltonian $\hat{\mathcal{H}}_{\rm res}$ (Eq.~\eqref{ising1}) for an optimal duration, diagonal expectation values (i.e., quantum features) are fed into a classical linear regression algorithm to reconstruct past inputs. \textbf{(b)} The model can undergo a spectral transition at a finite critical measurements rate $\gamma^{\rm c}$, where the spectrum becomes complex. Remarkably, this transition, which is due to a spontaneous symmetry breaking, coincides with a dynamical learnability transition at the exceptional point. 
}
\label{fig:schematic}
\end{figure}

In the present work, we address these dual objectives through the lens of NH Hamiltonians; a setting in which memory, tunability, and computational richness can be parsimoniously realized. Specifically, we study a generic NH many-body spin system, with an effective interpretation as a continuously measured system, postselected to retain only \textit{no-jump} trajectories~\cite{ashida2020non,RevModPhys.70.101,PhysRevLett.125.260601, PhysRevX.4.041001}. This picture corresponds formally to a complexification of the Hamiltonian spectrum, and is closely related to both imaginary-time evolution and state preparation protocols~\cite{PRXQuantum.2.010342,PhysRevLett.98.140506,PhysRevLett.131.110602,PRXQuantum.3.010320,motta2020determining,PhysRevLett.125.010501,wu2019observation,li2019observation, PhysRevLett.129.070401,chen2021quantum,PhysRevLett.127.020504,PRXQuantum.4.010324}. 

Extending dynamics via a NH Hamiltonian admits a number of exotic phenomena, including exceptional points, (spontaneous) symmetry-breaking~\cite{ozdemir2019parity,ding2022non} and measurement-induced phase transitions~\cite{fisher2023random}. Crucially, these systems may undergo dynamical purification and entanglement transitions~\cite{PhysRevLett.126.170503,PhysRevX.13.021007, PhysRevB.108.134305,PhysRevB.107.L020403,PhysRevB.103.224210,PhysRevB.107.L220201, PhysRevB.101.184201,PRXQuantum.5.030329}, and violate the bounds for spatiotemporal propagation of information  obeyed in unitary systems~\cite{PhysRevLett.124.136802, PhysRevLett.120.185301, PhysRevLett.125.260601, PhysRevB.110.094307}. Furthermore, in some disordered models, local randomness suppresses NH effects, leading to localization and a fully real spectrum in the thermodynamic limit~\cite{PhysRevLett.123.090603,mak2024statics,PhysRevA.108.043301}. The status of thermalization in NH systems also remains to be settled, with evidence both for violations of the eiegnstate thermalization hypothesis ~\cite{deutsch2018eigenstate} and opportunities for its  extension~\cite{cipolloni2024non,roy2023unveiling}. Remarkably, NH traits such as complex spectra and non-orthogonal states have been leveraged to enhance the performance of classical recurrent networks~\cite{kerg2019non,ganguli2008memory} and, more recently, quantum neural networks~\cite{sannia2024skin}. Furthermore, contemporary proposals for quantum simulation of nonlinear dynamics rely on NH Hamiltonians \cite{NOVIKAU2025109498}.

In sum, the present research landscape presents compelling evidence that NH systems offer additional dynamical complexity beyond the limitations of the traditional Hermitian paradigm of unitary quantum dynamics, with direct application to temporal quantum information processing. Here we make this explicit, arguing that NH systems are an ideal setting in which to instantiate QRC and QML. To do so, we first demonstrate in Sec.~\ref{sec:model} that exceptional points, which characterize distinct dynamical phases, may be controlled by the strength of local disorder and quantum interactions. These results are applied to the context of QRC in Sec.~\ref{sec:results}, where show that the real-to-complex spectral transition marked by the first exceptional point \textit{also} corresponds to a sudden change in the system's memory and state distinguishability properties. As Fig.~\ref{fig:schematic} illustrates, we find that the spectral transition induces a transition in learning capacity when the system is used as a reservoir. This alignment in physical and informational regimes (and their critical point) means there exists a controllable ``learning transition" for NH Hamiltonians. We further show that the learning performance is determined by the system's sensitivity to NH perturbations, which is itself determined by the presence of disorder and interactions. We conclude in Sec.~\ref{sec:discussion} by outlining the broader implications of this work, and its potential future development.

\section{Model \label{sec:model}}
\subsection{Controlling spectral properties of non-Hermitian Hamiltonians}

Guided by the need for tunable quantum memory and nonunitary dynamics in temporal learning processes, we introduce a NH spin Hamiltonian that serves as our computing reservoir:
\begin{equation}
    \hat{\mathcal{H}}_{\rm res}=\sum_{l,m,\alpha}J^{\alpha}_{lm}\,\hat{\sigma}^\alpha_l \hat{\sigma}^\alpha_m + \sum_{l,\alpha}h_l^{\alpha}\hat{\sigma}^{\alpha}_l - \frac{i\gamma}{2}\sum_l \hat{\sigma}^{y}_l,
    \label{ising1}
\end{equation}
where $\hat{\sigma}^{\alpha}_l$ denotes the $\alpha$ component of a Pauli spin-$1/2$ operator at location $l$ and $\alpha\in\{x,z\}$. Additionally, we define $h^{\alpha}_{l}=h^{\alpha}+\varepsilon^{\alpha}_l$, where $\varepsilon^{\alpha}_l\in [-\Delta^{\alpha},\Delta^{\alpha}]$ is a set of independent random variables. Following our recent work \cite{ivaki2024quantum}, the spins are arranged on vertices of \textit{random regular graphs}, with each vertex having exactly $k$ random edges. The coupling is defined as $J^{\alpha}_{lm}=J^{\alpha}A_{lm}$, where $A_{lm} \in \{0,1\}$ is an element of the adjacency matrix of graphs with $N$ spins. In what follows we set $(J^z, h^x, h^z)=(1,1,0)$, $\Delta^z=1$, and also fix $(N,k)=(8,4)$. All scales are given in terms of $J^z$ and the Hamiltonian $\hat{\mathcal{H}}_{\rm res}$ remains real irrespective of the parameters.
To explore the learnability properties of the model and map out the parameter space for optimization, here we mainly focus on the interplay of local disorder $\Delta^x$, the strength of NH perturbation $\gamma$, and the interaction $J^x$. Our results and conclusions remain consistent regardless of the specific initial parameters chosen, provided the general chaotic characteristics--such as spectral statistics and eigenstate thermalization--and the presence or absence of desired symmetries are maintained. 

\begin{figure}[t!]
\centering
\includegraphics[width=0.999\linewidth]{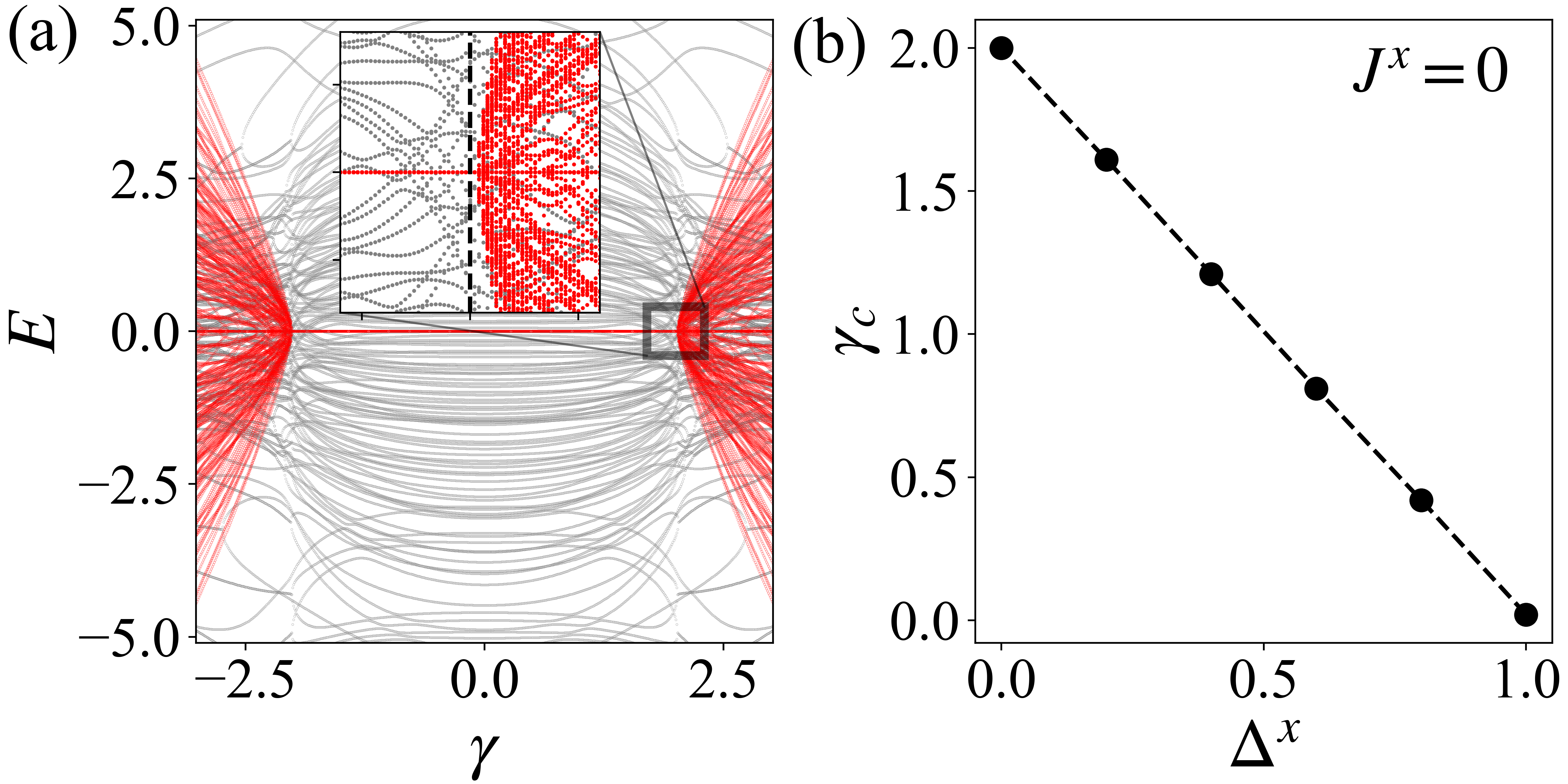}
\includegraphics[width=0.999\linewidth]{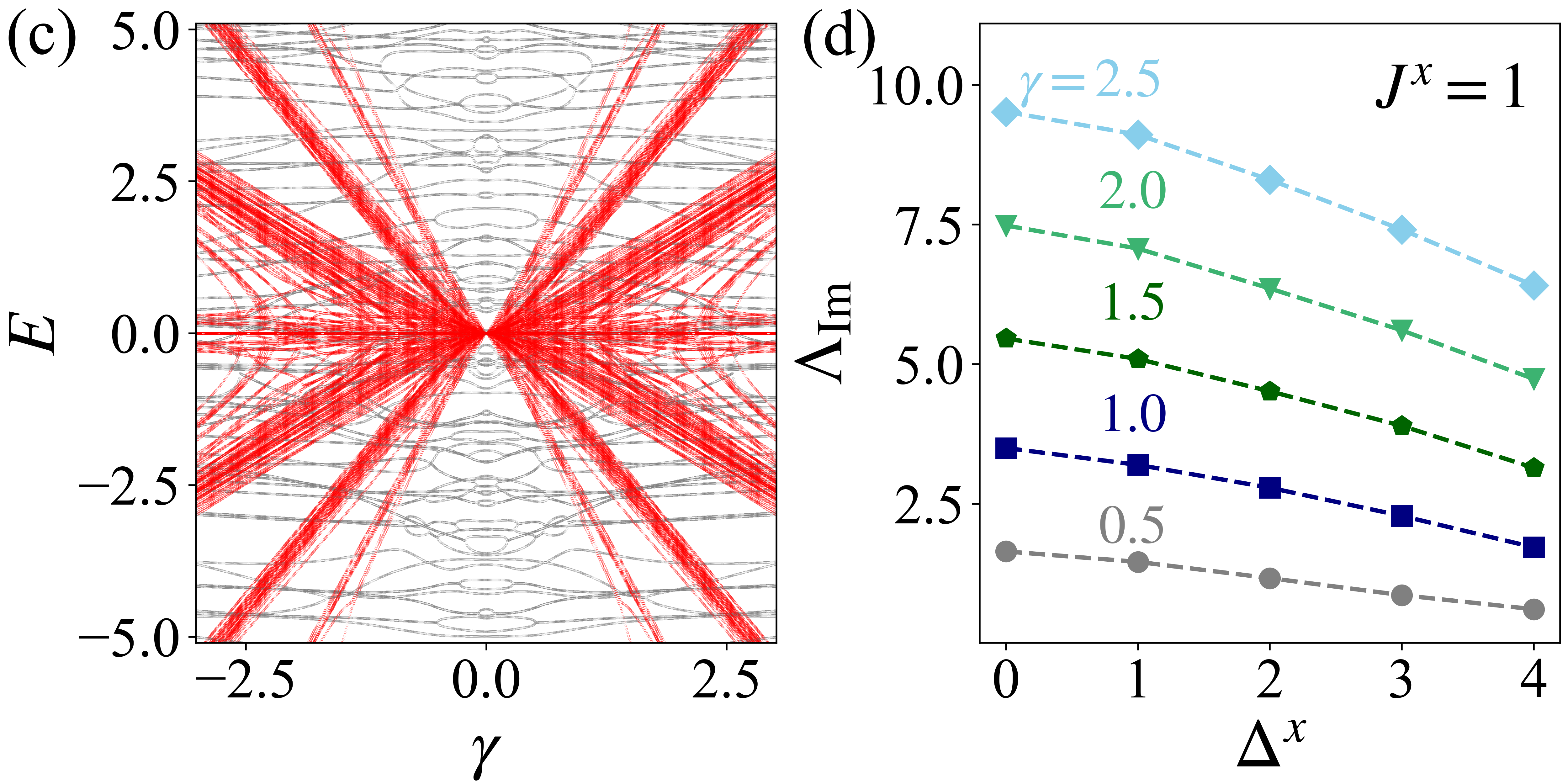}

\caption{ \textbf{Spectral properties.} \textbf{(a)} Spectrum of the model Eq.~\eqref{ising1} with $(J^x, \Delta^x)=(0,0)$, hosting a real-to-complex transition at $\gamma^c\approx2$. Grey and red lines denote the real and imaginary parts of the eigenenergies, respectively, and $\gamma^c$ is marked by the dashed line in the inset. \textbf{(b)} Appearance of first complex eigenergies as a function of the disorder strength $\Delta^x$, identified by pinpointing the critical value $\gamma^c$ of NH perturbation. This linear relation marks a learnability line and is expected to be exact in the limit $N\to\infty$. \textbf{(c)} Spectrum of the model for a typical random realization with $(J^x, \Delta^x)=(1,0)$. In this case $\gamma^c\approx0$ and most imaginary parts grow linearly with increasing $\gamma$, due to the linear dependence of the $\hat{\mathcal{H}}_{\rm res}$ on the NH parameter.
\textbf{(d)} Averaged maximum value $\Lambda_{\rm Im}$ of imaginary parts of eigenenergies, plotted for different values of $\gamma$. In this limit $\Lambda_{\rm Im}$ generally appears to be a decreasing function of $\Delta^x$. Each data point is a result of averaging over $50-100$ independent random realizations of disorder and random graphs. }
\label{fig:spectral_prop}
\end{figure}

It is useful to outline key properties of the model Hamiltonian Eq.~\eqref{ising1} in the context of the parity-time (\(\mathcal{PT}\)) symmetry-breaking transition~\cite{RevModPhys.96.045002,el2018non}. Within this framework, one finds a class of \textit{physical} systems that possess real spectra even though they are not Hermitian. Systems with balanced gain and loss provide a notable example in this regard~\cite{ashida2017parity,regensburger2012parity}. The presence of this symmetry, or more broadly, \textit{pseudo-Hermiticity}~\cite{mostafazadeh2010pseudo}, can signal a completely real spectrum for complex matrices. For a pseudo-Hermitian operator $\hat{\cal{H}}$ we can always write $\hat{\cal{H}}^{\dagger}\hat{\xi}=\hat{\xi}\hat{\cal{H}}$, where $\hat{\xi}$ is Hermitian and invertible, but not necessarily positive-definite. In other words, $\cal{PT}$ symmetry implies the existence of a pseudo-Hermitian structure, which underpins the reality of the spectrum~\cite{mostafazadeh2001pseudo}. A system is called \(\mathcal{PT}\)-unbroken or \textit{quasi-Hermitian} when \textit{all} of its eigenstates respect the symmetry, resulting in an \textit{entirely} real spectrum. This also implies that in this case $\hat{\cal{H}}$ can be transformed into a Hermitian operator via a Hermitian and positive-definite (but not necessarily with a bounded inverse) similarity transformation. Conversely, if some eigenstates do not, the symmetry is spontaneously broken, and the eigenvalues emerge as complex conjugate pairs~\cite{ashida2020non}. The symmetry-breaking transition, which notably takes place even in finite-size systems~\cite{bender2007making}, is typically signaled by the appearance of highly degenerate \textit{exceptional points }as the non-Hermiticity strength \(\gamma\) is tuned, leading to linearly dependent eigenvectors and distinctive physical effects~\cite{ashida2020non,bender2007making,el2018non,takasu2020pt,PhysRevB.110.014441}. These properties will reveal their importance in the present context when we discuss state distinguishability later.

In the limit $(J^x,\Delta^x) = (0,0)$, the Hermitian part of $\hat{\mathcal{H}}_{\rm res}$ describes a (non-integrable) transverse-field Ising model. When the dissipative $\hat{\sigma}^y$ term is introduced, spontaneous $\mathcal{PT}$ symmetry breaking may occur in the many-body spectrum near a critical strength $\gamma^{\rm c}$, as demonstrated in earlier studies for one-dimensional spin chains~\cite{PhysRevLett.126.170503,PhysRevB.110.094307,PhysRevB.108.134305}. Notice also that in the $(J^x,\Delta^x) = (0,0)$ limit the combination $\hat{\sigma}^{\pm} := h^x \hat{\sigma}^x \pm \frac{i\tilde{\gamma}}{2}\,\hat{\sigma}^y$ for $\tilde{\gamma}=2h^x$ renders $\hat{\mathcal{H}}_{\rm res}$ either upper or lower triangular in the $z$-basis. In this limit, the degeneracy coincides with the first exceptional point in the many-body spectrum where eigenvalues coalesce and we have $|\gamma^{\rm c}|\approx\tilde{\gamma}$. Thus, here the terms “degenerate point” and “exceptional point” are used interchangeably. Note it is due to the presence of this symmetry that the degenerate point can be obtained via only a single tuning parameter~\cite{PhysRevB.99.041406,PhysRevB.99.041202}. For $\gamma < |\gamma^{\rm c}|$, all eigenvalues remain real, although the eigenvectors need not be orthogonal. In contrast, for $\gamma > |\gamma^{\rm c}|$, some or all eigenvalues appear in complex conjugate pairs (with the same degeneracy count), as illustrated in Fig.~\ref{fig:spectral_prop}(a).

These observations suggest the interesting possibility that both the local randomness $\Delta^x$ the spin interaction $J^x$ could be used to control the spectral properties of the Hamiltonian. At the single particle level, we can immediately conclude that in the presence of random local fields, the critical point of the spectral transition is locally modified to \(\gamma^{\rm c}_{l}\approx \pm 2\left(h^{x}+\varepsilon^{x}_l\right)\). This simple picture remains valid in our many-body model too, and as shown in Fig.~\ref{fig:spectral_prop}(b) for \(0\leq\Delta^x\leq h^x\) and $h^x=1$, the averaged global critical rate obeys the linear relation \(\overline{\gamma^{\rm c}}\propto 1-\Delta^x\), marking the threshold at which the lowest exceptional point appears. Intuitively, $\left|\gamma^c_{\rm first}\right|\approx\min_l[2\left|h^x+\varepsilon^x_l \right|]= 2\left|h^x-\Delta^x \right|$, where $\gamma^{c}_{\rm first}$ denotes the critical NH perturbation strength for when the first (lowest-threshold) exceptional point can emerge. In the limit \(J^x=0\), this threshold effectively defines a learning transition point, whose significance we address in detail later. As the interaction \(J^x\) is gradually increased, the system becomes more sensitive to NH perturbations. For instance, as illustrated in Fig.~\ref{fig:spectral_prop}(c) for \(J^x/J^z=1\), complex eigenvalues emerge for almost any \(\gamma\neq 0\), with their magnitude growing linearly with \(\gamma\). As explained in Appendix~\ref{appendixb} through a mapping to a bosonic system, this effect can be attributed to the presence of two-body pairing terms, which can allow for immediate complex energy shifts. This regime is particularly suitable for learning applications, as the effective reservoir Hamiltonian, \(\hat{\mathcal{H}}_{\rm res}\), can in principle generate a \textit{contractive} dynamical map even for arbitrarily small values of \(\gamma\)~\cite{PhysRevLett.119.190401}, enabling the universality of the reservoir over a broad parameter range. 

In Fig.~\ref{fig:spectral_prop}(d) we further show the maximum imaginary part of the eigenenergies, \(\Lambda_{\rm Im}=\max_{j}\left|\operatorname{Im}E_j\right|\), for different disorder strengths. Increasing randomness appears to counteract the effect of NH perturbations, as indicated by the suppression of \(\Lambda_{\rm Im}\) with larger \(\Delta^x\). This behavior is likely due to the renormalization of the NH parameter to smaller effective values by preventing the proliferation of degenerate points where eigenenergies coalesce~\cite{PhysRevLett.123.090603, mak2024statics}. Accordingly, the interplay between \(\gamma\) and \(\Delta^x\) can serve as an effective tuning parameter for learning applications for typical QRC systems. Note that for a fixed \(\gamma\) and relatively weak disorder, \(\Lambda_{\rm Im}\) may in certain models grow with increasing the system size \(N\)~\cite{PhysRevLett.123.090603}. The scaling behavior and the role of connectivity \(k\) in this regard merit further investigations.
% #######################################################################
% #######################################################################
\subsection{Nonunitary dynamics}
\label{sec:dynamics}
To set the stage for QML and QRC, we need to understand quantum dynamics under NH Hamiltonians. A common and useful approach to capture NH effects is to study the dynamics of quantum states under general measurements. In this way, one may define a collection of positive semi-definite operators from a positive operator-valued measure (POVM) that, unlike projective measurements, are not limited to orthogonal projectors~\cite{wiseman2009quantum}. Such general measurement processes can be implemented by performing a projective measurement on an enlarged state space, also known as Naimark's dilation~\cite{PhysRevLett.101.230404,holevo2019quantum}. Thus the system may interact with an ancillary system for some time before a projective measurement is performed on the ancilla~\cite{shen2025observation,martinez2023certification}. In a discrete picture where the system evolves as \( \hat{\rho}(t+\delta t) \approx \sum_i \hat{\mathbf{K}}_i (\delta t)\, \hat{\rho}(t) \, \hat{\mathbf{K}}_i^{\dagger}(\delta t) \), one can define two categories of POVM elements in terms of the Kraus (measurement) operators: \( \hat{\mathcal{M}}_0=\hat{\mathbf{K}}_0^{\dagger}\hat{\mathbf{K}}_0 \), \( \hat{\mathcal{M}}_l=\hat{\mathbf{K}}_l^{\dagger}\hat{\mathbf{K}}_l \), satisfying \( \hat{\mathcal{M}}_0 + \sum_l \hat{\mathcal{M}}_l = \hat{\mathbb{I}} \). The measurement operators are given by 
\begin{equation}
\hat{\mathbf{K}}_0 = \hat{\mathbb{I}} - i\hat{\mathcal{H}}_{\rm res} \delta t, \quad \hat{\mathbf{K}}_l = \sqrt{\gamma \delta t} \, \hat{\mathcal{L}}_l,
\end{equation}
where \( \hat{\mathcal{L}}_l \)'s are called the jump operators. The effective reservoir Hamiltonian \( \hat{\mathcal{H}}_{\rm res} \) is related to the jump operators through 
\begin{align}
\label{eq:H_res_jump}
\mathcal{\hat{H}}_{\rm res} = \hat{\mathcal{H}} - \frac{i\gamma}{2} \sum_l \hat{\mathcal{L}}_l^\dagger \hat{\mathcal{L}_l}, 
\end{align}
with $\hat{\mathcal{H}}$ generating the unitary part of the evolution. The probability of a quantum jump occurring is given by
\(p_l = \text{Tr} \left(  \hat{\rho}(t)\, \hat{\mathcal{M}}_l \right) \approx \gamma \delta t \, \text{Tr} \left(  \hat{\rho}(t)\, \hat{\mathcal{L}}_l^\dagger \hat{\mathcal{L}}_l\right)+\mathcal{O}(\delta t^2)
\), and the probability of no jump is $p_0=1-\sum_l p_l$. 

Alternatively, for a continuous description of the evolution at the level of a single measurement trajectory we employ a stochastic Schrödinger equation (SSE) for quantum trajectories~\cite{ashida2020non,jacobs2006straightforward}, expressed in the form of:
\begin{align}
\delta \hat{\rho}(t)&= -i \left( \hat{\mathcal{H}}_{\rm res} \hat{\rho}(t) - \hat{\rho}(t) \hat{\mathcal{H}}_{\rm res}^\dagger \right)\delta t + \gamma\sum_l {\langle\hat{\mathcal{L}}_l^\dagger \hat{\mathcal{L}}_l\rangle}\hat{\rho}(t)\,\delta t  \nonumber \\
&\quad + \sum_l \left( \frac{\hat{\mathcal{L}}_l \hat{\rho}(t) \hat{\mathcal{L}}^{\dagger}_l} {{\langle\hat{\mathcal{L}}_l^\dagger \hat{\mathcal{L}}_l\rangle}} - \hat{\rho}(t)\right) d{\cal N}_l(t),
\label{eq:SSE}
\end{align}
where \( d{\cal N}\) represents the stochastic noise associated with continuous measurements with an average \( \overline{d{\cal N}_l(t)}= \gamma \,\delta t \langle\hat{\mathcal{L}}_l^\dagger \hat{\mathcal{L}}_l\rangle \). Here, \( \langle \cdots\rangle \) denotes the (time-dependent) expectation value with respect to the current state of the system. Averaging over the measurement outcomes, the ensemble-averaged density matrix evolves according to the standard master equation for open quantum systems~\cite{breuer2002theory}. However, with postselection of trajectories without jumps, known as the \textit{no-click} or \textit{no-count} limit~\cite{ashida2020non, PhysRevB.107.L020403,PhysRevA.100.062131,roccati2022non}, the conditional evolution becomes deterministic. This amounts to setting $d{\cal N}_l(t)=0$ in Eq.~\eqref{eq:SSE}, corresponding to null measurement outcomes, resulting in a nonunitary evolution governed by an \textit{effective} NH Hamiltonian. The postselected dynamics (omitting the normalization term) is then given as $\delta \hat{\rho}(t)= -i\delta t \left( \mathcal{\hat{H}}_{\rm res} \hat{\rho}(t) - \hat{\rho}(t) \mathcal{\hat{H}}_{\rm res}^\dagger \right).$ In this picture, the Hamiltonian in Eq.~\eqref{ising1} describes a scenario where continuous weak measurement of local spins along the $y$-direction is performed on all qubits, followed by postselection of a specific measurement outcome. Equivalently, as shown in Appendix~\ref{appendixb}, we can interpret the nonunitary effect as an asymmetric creation or destruction of degrees of freedom due to an incoherent external drive. Here then, taking $\hat{\mathcal{U}}_{\rm res}(\delta t)=\exp{(-i \delta t\, \hat{\mathcal{H}}_{\rm res})}$ and defining $\Omega[\hat{\rho}(t)]:=\hat{\mathcal{U}}_{\rm res}(\delta t) \, \hat{\rho}(t) \,\hat{\mathcal{U}}_{\rm res}^{\dagger}(\delta t)$, we arrive at a familiar expression for evolution of density matrices under the reservoir Hamiltonian,
\begin{align}
\hat{\rho}(t+\delta t)=\frac{\Omega[\hat{\rho}(t)]} {{\rm Tr}\left[\Omega[\hat{\rho}(t)]\right]}.
\label{eq:ures_map}
\end{align}
By including the normalization condition, this describes a completely positive and trace-preserving map. The map is also \textit{nonunital} since $\Omega[\hat{\mathbb{I}}]\neq \hat{\mathbb{I}}$. We note that in the quasi-Hermitian ($\mathcal{PT}$-unbroken) phase, although the eigenvalues of $\hat{\mathcal{H}}_{\rm res}$ are all real, the normalization factor exhibits oscillatory behavior due to nonorthogonality of eigenstates. This behavior distinguishes the dynamics from the usual unitary case. 

Crucially, the quantum channel generated by $\mathcal{\hat{U}}_{\rm res}(\delta t)$ in its general form can represent a purifying quantum map; \textit{i.e.}, ${\rm{Tr}}[\hat{\rho}^2(t)]\to1$ for $t,N\to\infty$. To see this, note that, in case of non-degenerate $\mathcal{\hat{H}}_{\rm res}$, and by using $\mathcal{\hat{H}}_{\rm res}=\sum_j \,E_j\ket{R_j}\bra{L_j}$, we may write $\mathcal{\hat{U}}_{\rm res}(\delta t)=\sum_j\, e^{-i\delta tE_j}\ket{R_j}\bra{L_j},$ where $\ket{R_j}$ and $\bra{L_j}$ are the so-called right and left eigenvectors of $\mathcal{\hat{H}}_{\rm res}$, respectively. They are defined such that $\mathcal{\hat{H}}_{\rm res}\ket{R_j}=E_j\ket{R_j}$ and $\bra{L_j}\mathcal{\hat{H}}_{\rm res}=\bra{L_j}E_j$. The eigenvectors satisfy the biorthogonal relationship $\braket{L_i}{R_j}=\delta_{ij}$, and form a complete biorthonormal eigenbasis $\sum_j \ket{R_j}\bra{L_j}=\hat{\mathbb{I}}$~\cite{brody2013biorthogonal}. Without normalization, we can write

\begin{align}
\label{eq:rhodt_lr}
    \hat{\rho}(\delta t) = \sum_{kj} e^{-i\delta t\,(E_k-E_j^*)} \rho_{kj}\, |R_k\rangle  \langle R_j|
\end{align}
where $\rho_{kj}=\langle L_k| \hat{\rho}(0)|L_j\rangle$. Since in the symmetry-broken regime the eigenvalues $E$ are generally complex with differing imaginary parts, the evolution will project any initial state into that of the fastest growing or slowest decaying eigenspaces at sufficiently long times~\cite{roy2023unveiling, PhysRevLett.126.170503}. The form of the evolution given by Eq.~\eqref{eq:rhodt_lr} thus suggests that in an optimal regime the dynamical map generated by Eq.~\eqref{eq:ures_map} can be contractive. We present evidence for this in Sec.~\ref{sec:results}. This property is vital to learnability performance of the physical network and, more fundamentally, to satisfying the universality criteria in memory-dependent machine learning tasks~\cite{sannia2024dissipation,sannia2024skin,nokkala2021gaussian}. 

We conclude this section by noting the intrinsic challenge in realising NH dynamics in a controllable manner. Yet if the goal is not merely to observe the features of these systems in simulation, but to exploit them—for memory storage, dynamical learning, or state preparation—then a systematic and physically implementable methodology is required. In particular, it is necessary to determine how the essential features of NH evolution can be replicated using standard quantum operations. While methods such as Linear Combination of Hamiltonian Simulation (LCHS) \cite{PhysRevLett.131.150603} offer a formally efficient route to nonunitary dynamics, they rely on complex ancilla-based constructions and optimal state preparation routines that are fundamentally ill-suited to the constraints of NISQ-era hardware. An alternative, constructive approach can instead be developed using the framework of \textit{Quantum Dynamical Emulation} (QDE) \cite{QDE}, which employs ensembles of unitary evolutions—with analytically prescribed weightings—to emulate nonunitary behavior. While our focus in the present work is to demonstrate the utility of NH dynamics, a QDE-based emulation protocol capable of reconstructing this behaviour is sketched in App. \ref{sec:generating_non_hermitian}.
% #######################################################################
% #######################################################################
\subsection{Data encoding and learning}

We now turn to the application of the NH Hamiltonians to quantities relevant to QML and QRC. Unlike conventional quantum neural network approaches such as variational quantum algorithms~\cite{havlivcek2019supervised,cerezo2021variational}, QRC architectures do not require training of an extensive number of quantum parameters; instead, a fixed quantum evolution serves as a computational reservoir, and only a classical readout layer is optimized utilizing mostly simple linear regression techniques~\cite{mujal2021opportunities}. In order to encode classical inputs for temporal information processing, we consider the following encoding Hamiltonian \( \mathcal{\hat{H}}_{\rm enc} \), that feeds an input \( \theta_n \) to all qubits at step \( n \) as
\begin{equation}
    \hat{\mathcal{H}}_{{\rm enc},n} = \sum_l \left( \theta_n + \delta_{ln} \right) \hat{\sigma}_l^x.
    \label{kick}
\end{equation}
Here \( \theta_n \in \left[ 0,1 \right] \) and \( |\delta_{ln}| \ll 1\) represents a small step-dependent random encoding error at different locations. This formulation for encoding is natural to programmable quantum devices~\cite{yasuda2023quantum}. 
Defining a periodic evolution operator as $\hat{\mathcal{V}}_n := \hat{\mathcal{U}}_{\rm res} \, \hat{\mathcal{U}}_{{\rm enc},n},$ the state of the system after the $n$th input is given by $\hat{\rho}(nt) = \hat{\mathcal{V}}_n \cdots \hat{\mathcal{V}}_1 \, \hat{\rho}(0) \, \hat{\mathcal{V}}_1^\dagger \cdots \hat{\mathcal{V}}_n^\dagger$, where $\hat{\rho}(0)$ is a random mixed state. Here, $\hat{\mathcal{U}}_{{\rm enc},n}=\exp{(-it'\mathcal{\hat{H}}_{{\rm enc},n})}$ and we work in the limit $\theta t'\ll 1 $. 

For a successful information processing, the observables of the system, which are expectation values of spin operators measured in a suitable basis, must respond distinctively to the sequence of input‑dependent kicks of Eq.~\eqref{kick}. More precisely, observables must \textit{solely} be functions of a \textit{finite} number of past inputs, as permitted by the extent of the system's memory, and similar input history should result in a similar reservoir state~\cite{nokkala2021gaussian}. The optimal window of the evolution under $\hat{\mathcal{U}}_{\rm res}$ yielding the best learning outcome is determined by the \textit{learning time} $J^zt_{\rm res}$, which, as we demonstrate later, is influenced by the dynamical phases of the system. For optimization and training purposes, we mainly consider diagonal local observables $\langle\hat{\sigma}_l^z(nt)\rangle={{\rm Tr}\left[\hat{\rho}(nt)\,\hat{\sigma}_l^z\right]}$ and correlation operators $\langle\hat{\sigma}_l^z(nt)\,\hat{\sigma}_m^z(nt)\rangle_{l\neq m}$ at the end of each encoding plus evolution cycle as time-dependent features. Note from Eq.~\eqref{eq:rhodt_lr} one can easily see that observables can be formally calculated solely in terms of the right eigenvectors~\cite{roy2023unveiling}. The (real parts of) these temporal features are then combined and used for linear regression and supervised learning to find an optimal set of weights for prediction of unseen data~\cite{mehta2019high} (see Appendix \ref{appendixa} for more details). 

As a memory capacity metric, we calculate the Pearson factor $\mathcal{C}_{n}$, a measure of information retrieval 
\begin{align}
\mathcal{C}_{n}:=\frac{{\texttt{cov}}^2(y_n,\tilde{y}_n)}{{\texttt{var}} (y_n) \, {\texttt{var}} (\tilde{y}_{n})},
\end{align}
where $y_n$ is the learning target and $\tilde{y}_n$ is the prediction at step $n$, and ${\texttt{cov}}$ and ${\texttt{var}}$ denote covariance and variance, respectively. For the accuracy metric we consider a normalized root mean-squared errors defined as
\begin{align}
 {\rm NRMSE}:=\sqrt{\frac{\sum_{n}^{N_{\theta}}(y_n-\tilde{y}_n)^2}{N_{\theta}\,{\texttt{var}} (\mathbf{y})}},
\end{align}
where $N_{\theta}$ is the length of the input series \(\mathbf{y}\). In what follows, we will see that the described systems with rich dynamical properties are ideally suited for quantum reservoir computing purposes and temporal data processing. 

It is worth noting that the degree of nonlinearity of the input-output map is generally determined by a combination of chosen set of observables, the form of input encoding, and the normalized (nonunitary) quantum map describing the evolution ~\cite{mujal2021analytical,govia2022nonlinear}. To summarize, in our setup first a random input is encoded via local rotations of qubits around the \( x \)-axis, then the reservoir undergoes a (postselected) nonunitary evolution, and subsequently features are collected in the \( z \)-basis for linear regression. This configuration is expected to introduce a nonlinear (trigonometric) input-output relation when considering the evolution of observables, as outlined in Appendix~\ref{appendixc}.
% #######################################################################
% #######################################################################
\section{Application to quantum machine learning \label{sec:results}}
\subsection{Distinguishability}

A necessary (but not sufficient) condition for the system to admit \textit{finite memory} and for computations to be independent of initial conditions is to fulfill the \textit{convergence} (echo-state) property~\cite{chen2019learning,nokkala2021gaussian}. To address this in our setting, we analyze the quantum map generated by Eq.~\eqref{ising1}, demonstrating that the above property can only be satisfied in the $\mathcal{PT}$-broken phase. A suitable measure in this regard is to calculate the trace distance $\mathcal{D}$ defined as $\mathcal{D}(\hat{\rho}, \hat{\rho}') = \frac{1}{2} \text{Tr} \left| \hat{\rho} - \hat{\rho}' \right|
$~\cite{PhysRevA.71.062310}. Here, $\hat{\rho}$ and $\hat{\rho}'$ are two random initial mixed-states and $\left|\hat{\rho}\right|=\sqrt{\hat{\rho}^{\dagger}\hat{\rho}}$. Operationally, given an optimal measurement strategy, the trace distance represents the maximal probabilistic advantage $p_{\rm adv}$ that can be gained $2p_{\rm adv}=1+\mathcal{D}(\rho,\rho')$ in distinguishing two random and equally probable density matrices. Note that the distance remains invariant under a unitary transformation of the states.

\begin{figure}[t!]
\centering
\includegraphics[width=0.999\linewidth]{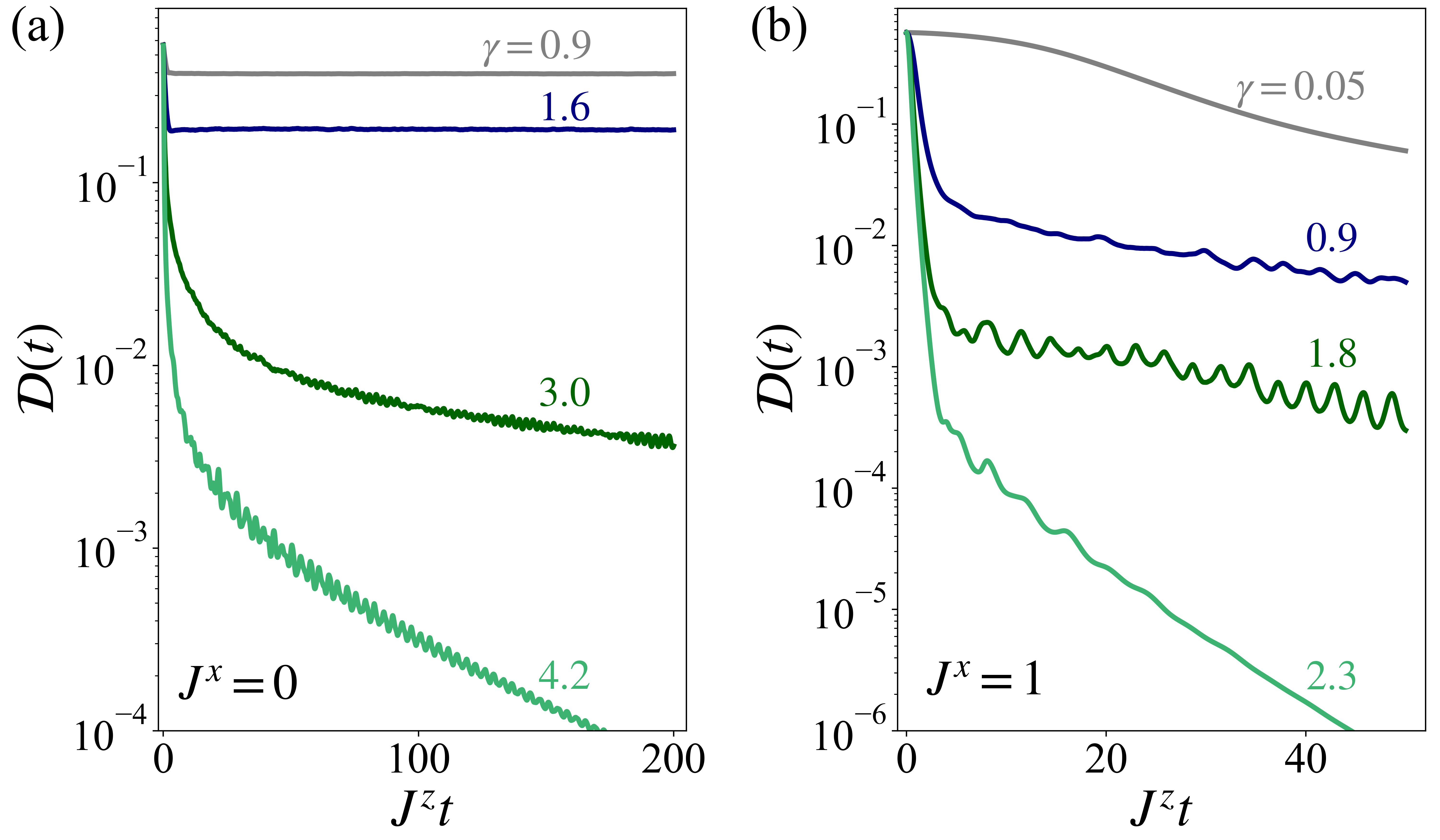}
\caption{\textbf{Dynamics of state distinguishability.} \textbf{(a)} Trace distance $\mathcal{D}(t)$ between two random mixed-states for $(J^x,\Delta^x,\delta t)=(0,0,0.05)$. In this case $\mathcal{D}(t)$ remains constant upon averaging for $\gamma<2$, and decays exponentially otherwise for $\gamma t\to\infty$. \textbf{(b)} Same for $(J^x,\Delta^x,\delta t)=(1,0, 0.01)$, where asymptotically there is a unidirectional flow to environment, and at long times we have $\dot{\mathcal{D}}(t)<0$ for any $\gamma>0$. The initial distance on average is $\mathcal{D}\left(\hat{\rho}(0), \hat{\rho}'(0)\right)\approx0.7$.
}
\label{fig_distance}
\end{figure}

In Fig.~\ref{fig_distance}(a), we plot the distinguishability between two random mixed states under the $\hat{\mathcal{H}}_{\rm res}$ for different strength of measurements $\gamma$ in the limit $(J^x,\Delta^x)=(0,0)$. We find that for $\gamma<\vert \gamma^{\rm c}\vert \approx 2$, where the spectrum is completely real, the distance remains constant. However, beyond $\vert \gamma^{\rm c}\vert$ it decays exponentially as $\mathcal{D}(t) \sim \exp{(-t/\zeta)}$ for $\gamma t\to \infty$, where $\zeta\propto \left[\Lambda_{\rm Im}\right]^{-1}\propto \gamma^{-1} $ determines the characteristic timescale of relaxation (or instability). Far from the critical point we roughly get $\delta \log\mathcal{{D}} \propto  -\gamma\delta  t$. Furthermore, Fig.~\ref{fig_distance}(b) depicts the dynamics of distinguishability for finite interaction $J^x/J^z=1$, where there is no finite symmetry-breaking point, and for an infinitesimally small NH perturbation there are some eigenenergies that quickly develop an imaginary part, as previously shown in subsection~\ref{sec:model}. Thus, in principle a finite system can be efficiently purified with a $\gamma$-dependent rate for any $\gamma\neq0$. These are the first main results here and indicate that the critical (or exceptional) point, whose location can be tuned by randomness and interactions, serves as a threshold for learning.

To carry out memory-dependent learning processes successfully, the dynamical map should be \textit{asymptotically Markovian}, in the sense that the rate of change $\mathcal{\dot{D}}\equiv\delta \left[ \mathcal{D}\left( \hat{\rho}(t), \hat{\rho}'(t)\right) \right]/\delta t $ must satisfy $ \dot{\mathcal{D}} < 0$~\cite{RevModPhys.89.015001, PhysRevLett.103.210401}. In the context of $\mathcal{PT}$-symmetric NH quantum systems, the above property strictly holds only when symmetry is broken and the reservoir features a complex spectrum~\cite{PhysRevLett.119.190401}. Since quasi-Hermitian Hamiltonians admit a completely real spectrum and can be mapped to Hermitian systems (via a similarity transformation), they generate a unitary-like evolution under a refined inner product space. Due to this they lack the dynamical properties that facilitate the decay of the averaged state distinguishability at long times and, consequently, the quasi-Hermitian systems on their own are generally less practical for the reservoir computing tasks of interest. As shown in Appendix~\ref{appendixd} for a two-level system, in this regime measures such as purity, entanglement, and trace distance may exhibit persistent $\gamma$-dependent oscillatory behavior, which can be interpreted as a (perpetual) full or partial recovery of dissipated information~\cite{PhysRevLett.119.190401, PhysRevLett.124.230402,PhysRevLett.123.230401}. In this view, $\gamma$ can control the degree of non-Markovianity of a NH model, potentially enhancing memory in specific tasks~\cite{sannia2025non}. 

In the next subsection we will see a direct manifestation of the above results in our model for a concrete learning task, tying the spectral properties of NH systems directly to quantum learning performance.

% #######################################################################
% #######################################################################
\subsection{Learnability and memory transitions}

The hallmark of a powerful QRC architecture is its ability to accurately recall past inputs over a finite, but sufficiently extended, time window. To quantify this, we first assess linear learnability in a minimal temporal task, where the learning target, $y_{n,\tau}$, is a delayed restoration of past inputs $\theta_{n-\tau}$. Formally, 
\begin{align}
y_{n,\tau}:=\theta_{n-\tau}. 
\end{align}
We define the total memory capacity as ${\mathcal{C}}_{\rm T}= \sum^{\tau_{\rm max}}_{\tau} \sum^{N_{\theta}}_{n}\mathcal{C}_{n,\tau}$, where $\mathcal{C}_{n,\tau}$ is the correlation coefficient for $y_{n,\tau}$. Here $\tau$ is an integer specifying the \textit{delay time} and in our calculations we always take $\tau>0$. Since $\mathcal{C}_{n,\tau}\to 0$ when $\tau\to \infty$, for a given maximum delay time $\tau_{\rm max}\geq1 $ we can normalize as $\overline{\mathcal{C}_{\rm T}}=\mathcal{C}_{\rm T}/\tau_{\rm max}$. The total normalized memory $\overline{\mathcal{C}_{\rm T}}$ is bounded between unity and zero, indicating complete or vanishing correlation to previous inputs, respectively. 
\begin{figure*}[t!]
\centering
\includegraphics[width=0.95\linewidth]{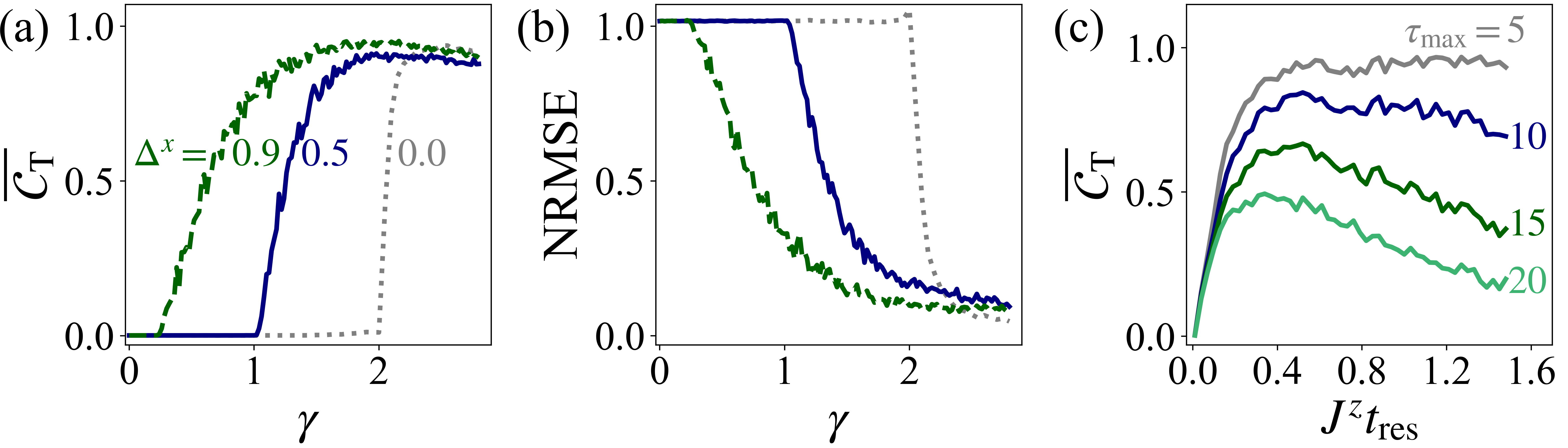}
\includegraphics[width=0.95\linewidth]{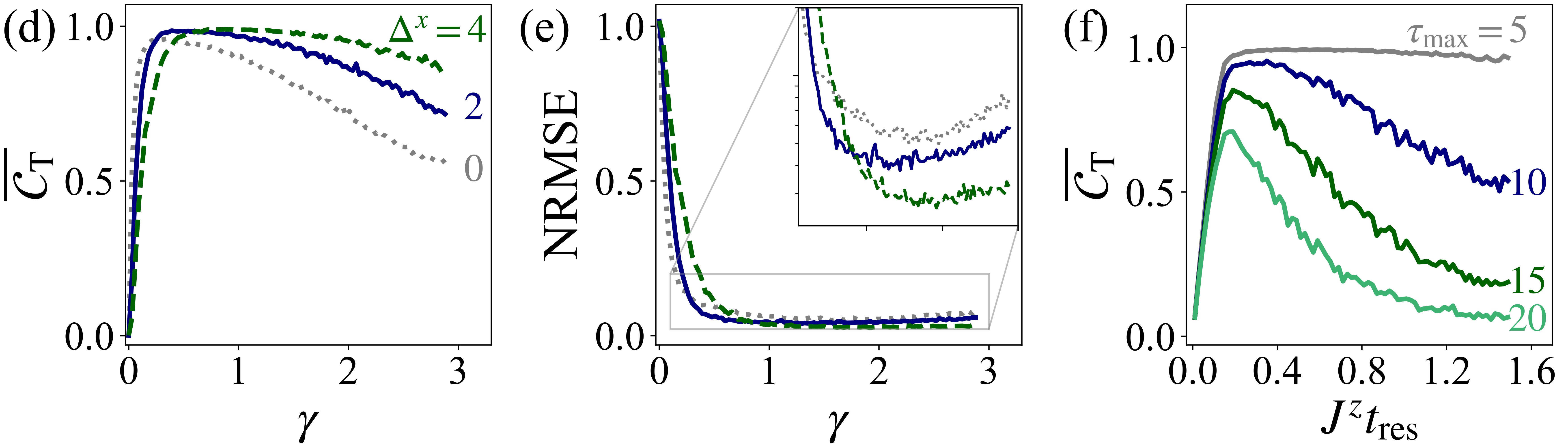}
\caption{ \textbf{Linear temporal learning.}
\textbf{(a)} Normalized averaged total memory capacity $\overline{\mathcal{C}_{\rm T}}$ as a function of measurements rate $\gamma$, plotted for $(J^x, \tau_{\rm max}, Jt_{\rm res})=(0,10,0.4)$. \textbf{(b)} NRMSE for $(J^x, \tau, Jt_{\rm res})=(0,3,0.4)$. The critical point of learning transitions coincides with the points where the spectrum of $\hat{\mathcal{H}}_{\rm res}$ becomes complex. \textbf{(c)} $\overline{\mathcal{C}_T}$ as a function of the learning time $Jt_{\rm res}$, plotted for different values of maximum delay $\tau_{\rm max}$, with $(J^x,\Delta^x,\gamma)=(0,0.5,1.8)$. For a fixed $\gamma,\Delta^x$, there is an optimal interval at short times where the memory is maximized. \textbf{(d)} $\overline{\mathcal{C}_{\rm T}}$ for $(J^x, \tau_{\rm max}, Jt_{\rm res})=(1,10,0.25)$. \textbf{(e)} NRMSE for $(J^x, \tau, Jt_{\rm res})=(1,3,0.25)$. Inset displays the errors in the zoomed-in region on a logarithmic scale. \textbf{(f)} $\overline{\mathcal{C}_{\rm T}}$ as a function of $Jt_{\rm res}$ for different values of $\tau_{\rm max}$, with $(J^x,\Delta^x,\gamma)=(1,2,0.4)$. Compared to the case $J^x=0$, with interactions one might reach a maximal learning performance at shorter times and with weaker nonunitary effects, alleviating the computational burden of practical simulations.
}
\label{fig:panel_memory_accuracy}
\end{figure*}

We first address the behavior of memory capacity for $J^x=0$ in the vicinity of the spontaneous $\mathcal{PT}$ symmetry-breaking transition in the many-body spectrum at $|\gamma^c|\approx2$. Remarkably, following the behavior of distinguishability discussed in the previous section, this transition corresponds to an abrupt learning transition, as seen in Figs.~\ref{fig:panel_memory_accuracy}(a)-(b). Moreover, as shown before, in this limit local disorder \( \Delta^x \) shifts the real-to-complex transition to lower critical values on average, thereby enabling finite memory at smaller NH perturbations. For an optimal time $J^zt_{\rm res}$, the learning threshold is a linear function of the disorder strength, matching the critical value of symmetry breaking. Figure~\ref{fig:panel_memory_accuracy}(c) also depicts the total memory capacity as a function of learning time $J^zt_{\rm res}$ for different values of maximum delay $\tau_{\rm max}$. For rather large delays the reservoir memory peaks at short times, which is practically advantageous.

For the case of finite interaction $J^x=1$, shown in Figs.~\ref{fig:panel_memory_accuracy}(d)-(f), the learning performance improves markedly even for very small values of $\gamma$. At stronger NH perturbation strengths, the learning performance again degrades, due to fast decay of the stored information. Randomness in this limit enables direct control over the system's sensitivity to nonunitary effects. In this regime and for a fixed network size, disorder mitigates the adverse effects of strong dissipation, supporting improved memory retention and learning accuracy, as evidenced also by the decreasing behavior of \(\Lambda_{\rm Im}\) when increasing disorder, displayed in Fig.~\ref{fig:spectral_prop}(d). Apart from this, as established in earlier studies~\cite{ivaki2024quantum,xia2022reservoir,Martinez-Pena2021,sannia2024skin}, incorporating controlled randomness can generally enhance the expressivity of a dynamical system through regulating the chaotic behavior, leading to a broader entropic distribution for features and allowing the system to represent more complex functions. 

\begin{figure}[t!]
\centering
\includegraphics[width=0.999\linewidth]{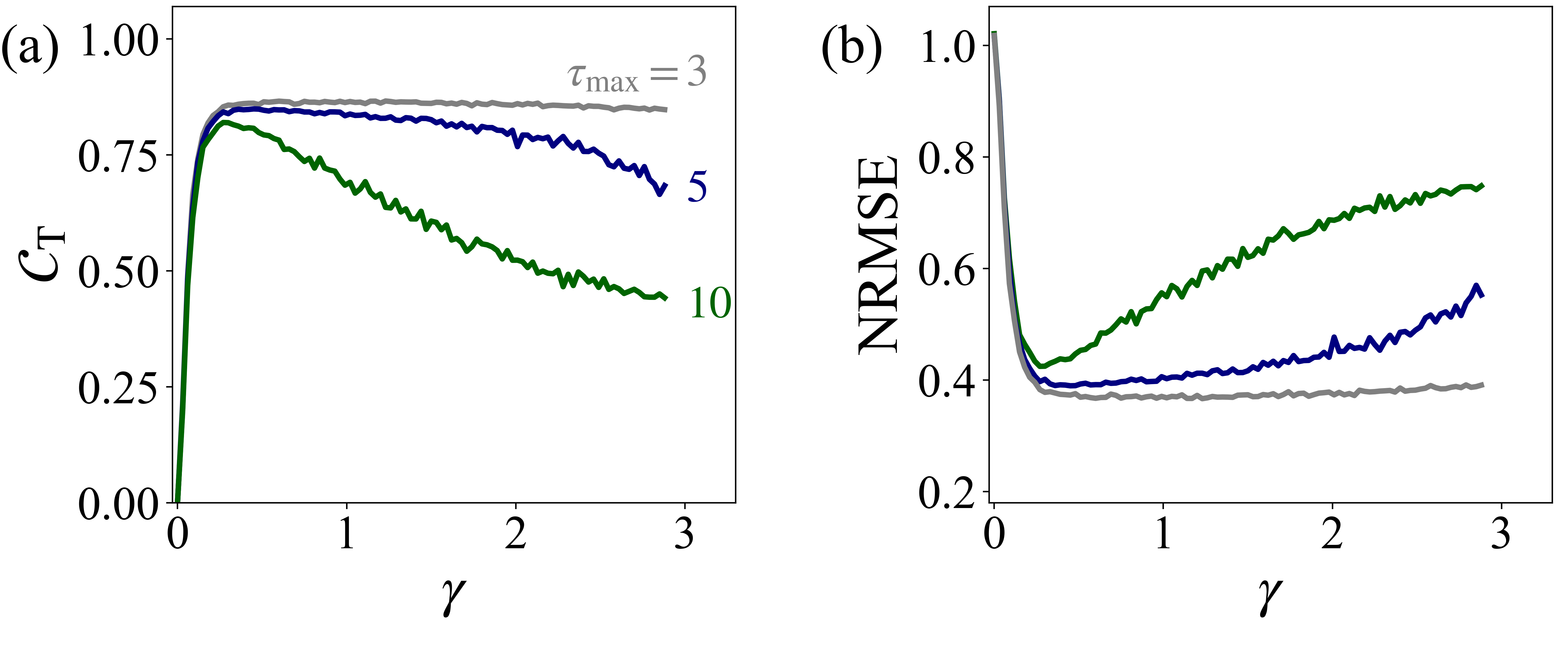}
\caption{ \textbf{Nonlinear temporal learning.} \textbf{(a)} Total memory capacity ${\mathcal{C}_{\rm T}}$ and \textbf{(b)} NRMSE as a function of the strength of non-Hermiticity $\gamma$, plotted for $(J^x,\Delta^x,Jt_{\rm res})=(1,2,0.3)$ and different degree of nonlinearity $\tau_{\rm max}=3,5,10$. This result establishes the ability of the NH system to successfully carry out a nonlinear temporal task in a generic parameter regime where NH effects are relatively weak.}
\label{fig:narma}
\end{figure}

We further examine the capability of the reservoir on processing a \textit{nonlinear} time-dependent series. A standard benchmark in this regard is reconstructing a nonlinear auto-regressive moving average (NARMA) series, expressed as
\begin{equation}
\begin{split}
y_n &= 0.3\, y_{n-1} + 0.05\, y_{n-1} \sum_{\tau=1}^{\tau_{\rm max}} y_{n-\tau} \\
    &\quad + 1.5 \,\theta_{n-\tau_{\max}}\, \theta_{n-1} + 0.1,
\end{split}
\end{equation}
The NARMA task has been studied extensively in the context of both RC and QRC~\cite{PhysRevApplied.8.024030,chen2019learning}. The summation term $\sum_{\tau=1}^{\tau_{\rm max}} y_{n-\tau}$ introduces a nonlinear moving average dependence with degree $\tau_{\rm max}$ and the term $\theta_{n-\tau_{\rm max}} \,\theta_{n-1}$ represents a multiplicative interaction between past inputs. In this way, the target series $\mathbf{y}$ is generated recursively from the random inputs, which are uniformly sampled in the interval $\theta_n\in[0,0.2]$. For this task total memory simply is $\mathcal{C}_{\rm T}=\sum_n \mathcal{C}_{n}$. As shown in Fig.~\ref{fig:narma} for a fixed learning time, $J^zt_{\rm res}$, and disorder strength, $\Delta^x$, the reservoir's performance is qualitatively similar to the linear case and comparable to some previous results~\cite{chen2019learning}. 
% The discussed sudden transition in linear memory at $J^x=0$ also takes place for this nonlinear task with the same linear relationship with disorder $\Delta^x$ for the transition point (not shown).

From the results here one can conclude that, in our model, randomness and interactions not only boost the learning performance, but potentially reduce the complexity of emulating NH dynamics as well. This is simply because a lower threshold $\gamma^c$ for a successful learning weakens the influence of nonunitary effects and effectively lowers the cost of postprocessing. Naively, the postselection cost (of ancillary measurements) for each copy of the system scales as \( \sim \mathcal{O}\left(2^{N T g(\gamma)}\right) \), where \( T \) is the total simulation time and \( g(\gamma) \) is a continuous function of the measurement rate. Accordingly, reducing the measurement rate strongly suppresses the exponential growth of postselection cost in the spacetime volume $NT$. Moreover, since learning performance peaks at relatively short times, the total input processing time is also reduced, making learning highly efficient. Thus, for our purposes, the simulability can be optimized without sacrificing learning capacity by operating near--but not deep into--the symmetry-broken phase. More broadly, the computational power of NH systems for solving general classes of decision problems is contingent upon the specifics of their singular-value spectra and the structure of their unitary purifications~\cite{barch2025computational}.

% #######################################################################
% #######################################################################
\subsection{Entanglement as a learning resource}
\label{sec:entanglement}

The nonunitary dynamics of NH Hamiltonians beyond the critical point and its implication to system purification raises an interesting question on quantum correlations and entanglement. While a complete characterization of quantum correlations is beyond the scope of the present work, we can calculate  
the averaged logarithmic negativity $\mathcal{E}(t)=\log_2\lVert\hat{\rho}^{T_{\mathcal{S}}}(t)\rVert$  as a relevant measure \cite{PhysRevA.65.032314,PhysRevLett.95.090503}. Here $T_{\mathcal{S}}$ indicates partial transpose with respect to a \textit{random subsystem} $\mathcal{S}$ of size $N/2$ and $\lVert\cdot\rVert$ is the trace norm. When a general quantum state is entangled, its partial transpose can have negative eigenvalues. We initialize the system in $\hat{\rho}=\ket{\Psi}\bra{\Psi}$, where $\ket{\Psi}=\ket{+}^{\otimes N}$ and $\ket{+}=\frac{1}{\sqrt{2}}(\ket{0}+\ket{1})$ is an eigenstate of the $\hat{\sigma}^x$ operator. We note that since there is a hidden measurement postselection in the NH dynamics, the purity of the initial product state is preserved. 

\begin{figure}[t!]
\centering
\includegraphics[width=0.999\linewidth]{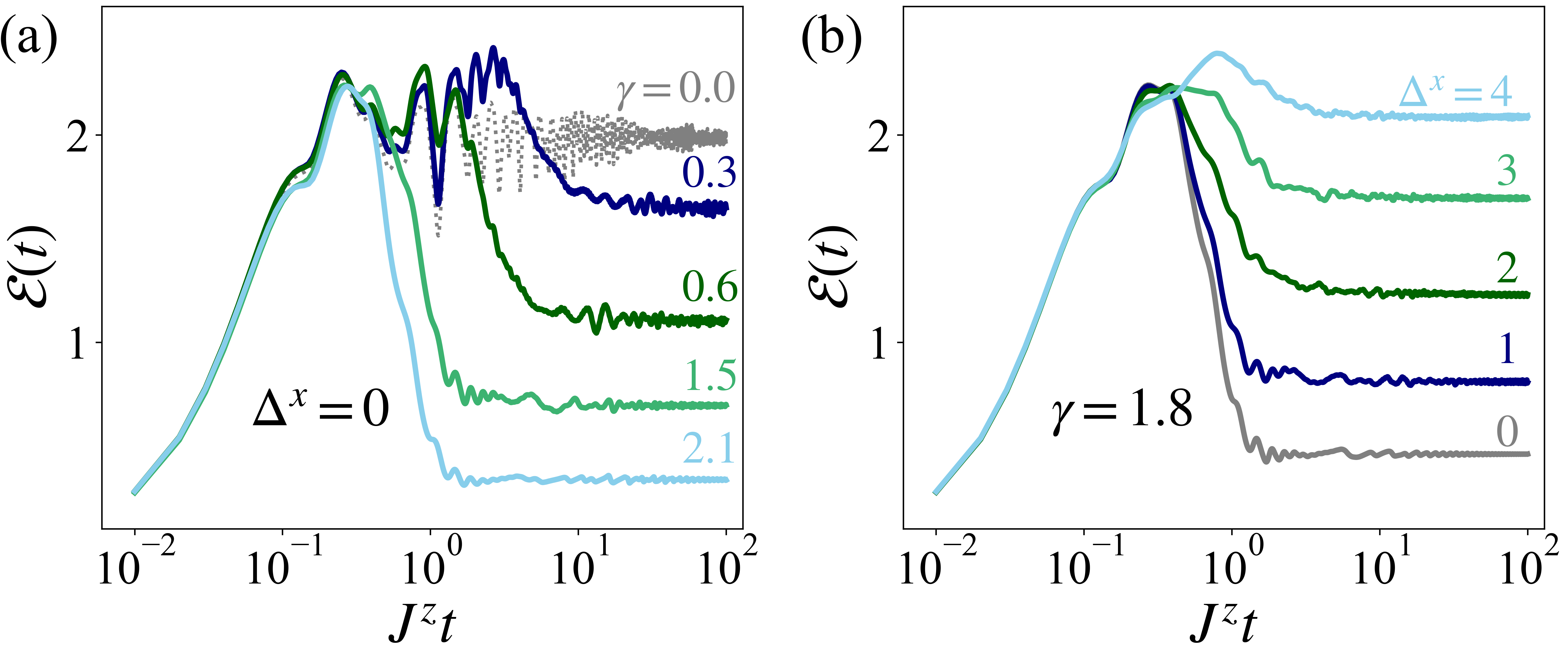}

\caption{ \textbf{Dynamics of quantum correlations.} Logarithmic negativity $\mathcal{E}(t)$ for \textbf{(a)} various strength of measurements rate $\gamma$, and \textbf{(b)} for various disorder strength $\Delta^x$. While the steady state correlation $\mathcal{E}(t)$ generally is a decreasing function of $\gamma$, in finite systems disorder on average might weaken the nonunitary effects, which introduces another degree of flexibility for optimizing the learning process. In this set $J^x=1$.}
\label{fig:negativity}
\end{figure}

As plotted in Fig.~\ref{fig:negativity}(a) for $J^x=1$, at finite measurement strength, $\gamma\neq0$, and without disorder, $\Delta^x=0$, the spin system may experience a dynamical entanglement transition to a (relatively) weakly-entangled subspace at long times, as established in previous studies~\cite{PhysRevB.108.134305,PhysRevLett.126.170503}. This transition in some cases can also be understood in terms of purification of maximally mixed states. Remarkably, as shown in Fig.~\ref{fig:negativity}(b), in our model moderate disorder seems to hinder the long-time evolution toward these weakly-entangled steady-states which also translates to a slower purification in typical finite-size quantum systems utilized for QRC purposes. In our setup, slower purification together with enhanced entanglement signify a stronger memory retention, which, as quantified in Figs.~\ref{fig:panel_memory_accuracy}(d)-(e), directly correlates with improved total memory capacity and learning accuracy. As such, local randomness in our model can serve as an optimization parameter to enhance and control the learning capabilities of a lossy computing reservoir. The thermodynamic behavior of the steady-state values and emergence of possible intermediate timescales can be explored in future works.
% #######################################################################
% #######################################################################
\section{SUMMARY AND DISCUSSION \label{sec:discussion}}
In this work we have investigated nonconventional dynamics as a means to realize nonconventional computing. In particular, we have sought to establish NH spin Hamiltonians as promising candidates for both QML and QRC. For the subclass of pseudo-Hermitian Hamiltonians in which exceptional points mark the onset of spectral complexification, we show that the location of the spectral transition can be controlled by local disorder and spin interactions. The significance and utility of this was assessed by its impact on quantum correlative figures of merit, where pronounced qualitative and quantitative changes were observed beyond a critical threshold. In the context of QRC, this was shown to correspond to an exploitable regime of enhanced prediction accuracy and memory capacity, and the identification of a ``learnability transition", where two dynamical phases with distinct learning capacities straddle a critical point. 

Notably, our results reveal that the parameter landscape of NH systems is highly optimizable; the learning transition can take place without a concomitant increase in the complexity required to simulate the underlying dynamics. In particular, learning is possible over a broad range of NH perturbation strengths and evolution timescales, meaning these quantities can be minimized while still maintaining the required degree of learning performance. Crucially, this suggests that controlled nonunitarity not only enhances computational capacity (via fading memory), but defines a tractable optimization surface well-suited to scalable implementation. 

The approach taken here aims to integrate physical principles into the design of QML architectures. By studying the link between physical and informational dynamics in the context of NH systems, we are able to frame learnability as a dynamical phase transition. Conceptually, this emphasizes the duality between material and computational processes, and the means by which they can be characterized simultaneously. Practically, it proposes a simple approach to reservoir design: a disordered many-body system, driven by NH perturbations into a learning phase. The framework presented is readily implementable via a variety of methodologies (see e.g. App. \ref{sec:generating_non_hermitian}), and aligns with emerging paradigms in neuromorphic quantum computing and quantum reservoir computing under noisy conditions~\cite{markovic2020quantum}. The learnability transition uncovered here could potentially be observed in quantum optical experiments where $\mathcal{PT}$ transitions have already been demonstrated~\cite{li2019observation,PhysRevLett.103.093902}, and platforms such as solid-state defects may also serve as promising candidates~\cite{PhysRevX.9.031045}. Future studies may build on these insights to develop quantum architectures that not only tolerate, but also actively harness noise as a computational resource~\cite{PRXQuantum.3.030101,PhysRevApplied.21.067001}.

\section{Data and code availability statement}

% All the relevant data shown in the figures here are available at xxxx.
All the relevant data presented here and the corresponding computer codes generating them are available upon a reasonable request.

\section{Acknowledgments}
We are grateful for valuable discussions with Alexander Balanov, Emil J. Bergholtz, Soon Hoe Lim, Jose L. Lado, Juan S. Totero Gongora, Viktor Ivády and Ali G. Moghaddam. We acknowledge the support of the UK Research and Innovation (UKRI) Horizon Europe guarantee scheme for the QRC-4-ESP project under grant no. 101129663 and EU Horizon Europe Quest project (no. 101156088). T.A-N. has been supported in part by the Academy of Finland through its QTF Center of Excellence program (project no. 312298). A.J.S. also graciously acknowledges support from the Magnus Ehrnrooth Foundation as well as the Finnish Ministry of Education and Culture through the Quantum Doctoral Education Pilot Program (QDOC VN/3137/2024-OKM-4) and the Research Council of Finland through the Finnish Quantum Flagship project (no. 358877, Aalto).
% #######################################################################
% #######################################################################
% #######################################################################
% #######################################################################
\appendix

\section{Details of numerical calculations}
\label{appendixa}

For training purposes, we have used the Ridge linear regression methods~\cite{mehta2019high} with the regularization strength set to $\lambda=10^{-2}\textendash 10^{-4}$, depending on the reservoir parameters and task. In ridge regression, we solve for the weight vector \( \mathbf{w} \) by minimizing the cost function
\begin{equation}
\min_{\mathbf{w}} \; \|\mathbf{Xw} - \mathbf{y}\|^2 + \lambda \|\mathbf{w}\|^2,
\end{equation}
which has the closed-form solution \({\mathbf{w}} = \left(\mathbf{X}^\top \mathbf{X} + \lambda \mathbb{I}\right)^{-1}\mathbf{X}^\top \mathbf{y}\). Here, \(\mathbf{X} \in \mathbb{R}^{T\times D}\) is the feature matrix, where \( T \) is the number of time steps (observations), and \( D \) is the number of features ( observables). Each row of \( \mathbf{X} \) corresponds to a time step, and each column to a specific feature, and  \(\mathbf{y}\) is the target vector, containing the input value for each observation. To further improve the stability of the regression, for each observable we utilize data standardization in training and testing stages as follows
\begin{equation}
    \tilde{\mathbf{x}}=\frac{\mathbf{x}-{\texttt{mean}}({\mathbf{x}_{\rm train})}}{{\texttt{stdv}}\, ( \mathbf{x}_{\rm train})}
\end{equation}
where $\mathbf{x}\in \mathbb{R}^{T}$ represents an array of a feature over time for training or test, and \texttt{mean} and \texttt{stdv} stand for the arithmetic mean and standard deviation, respectively.

The learning consists of three stages: washout, training, and testing, with the length of input series for each step set to $2\times10^3, 1.3\times10^3, 1.3\times10^3$, respectively. As discussed in the manuscript, the washout phase and in turn the effectiveness of learning are directly influenced by the dynamics of distinguishability and entanglement, which determine the speed of convergence and the nature correlations of the long-time steady-state, respectively. Therefore, an initial transient period for relatively long time steps is discarded to ensure applicability across a wide range of parameters. The expectation values of spin operators that are used to train the network are calculated for 100-200 independent combinations of disorder realizations, random graphs, and input sequences. In total for the number of $N=8$ spins we have collected 36 distinct features in the $z$-basis and solely at end of each input cycle. The reported results are then averaged over the outcome of the evaluation stage. While more sophisticated classical learning approaches may yield even better performance, here we have focused on the simplest approach possible to showcase the functionality of the reservoir.
% #######################################################################
% #######################################################################
\section{Mapping to hard-core bosons}
\label{appendixb}
To connect with previously studied models and gain insights into the role of different terms, we apply a standard local transformation that maps the spin-1/2 system studied in the main text to a hard-core boson representation. Using the identifications $\hat{\sigma}_l^- \leftrightarrow \hat{b}_l$ and $\hat{\sigma}_l^+ \leftrightarrow \hat{b}_l^\dagger$ with $\hat{\sigma}_l^y \leftrightarrow i(\hat{b}_l - \hat{b}_l^\dagger)$ and $\hat{\sigma}_l^z \leftrightarrow 2\hat{n}_l - 1$, we obtain the Hamiltonian
\begin{align}
    \hat{\cal{H}}_{\text{res}} = & \sum_{l,m} J_{lm}^x (\hat{b}_l^\dagger + \hat{b}_l)(\hat{b}_m^\dagger + \hat{b}_m) \nonumber \\
    & + \sum_{l,m} J_{lm}^z (2\hat{n}_l - 1)(2\hat{n}_m - 1) + \sum_{l} h_l^z (2\hat{n}_l - 1) \nonumber \\
    & + \sum_{l} \left[(h_l^x {-} \tfrac{\gamma}{2})\,\hat{b}_l^\dagger + (h_l^x {+} \tfrac{\gamma}{2})\,\hat{b}_l\right].
    \label{eq:app_hcb_final}
\end{align}
Here, the hard-core boson operators satisfy $(\hat{b}_l)^2 = (\hat{b}_l^\dagger)^2 = 0$ and $[\hat{b}_l, \hat{b}_m^\dagger] = (1 - 2\hat{n}_l)\,\delta_{lm}$, with $\hat{n}_l = \hat{b}_l^\dagger \hat{b}_l \in \{0,1\}$. In this language, the $J^x$ term includes both hopping and pairing contributions and explicitly breaks the $U(1)$ particle-number symmetry. The diagonal terms represent density-density interactions and onsite potentials, analogous to terms in Bose-Hubbard-type models~\cite{PhysRevLett.123.090603,mak2024statics}. The last term, 
$\hat{V}_{\rm NH} = \sum_{l} f_l(\gamma)\,\hat{b}_l^\dagger + f_l(-\gamma)\,\hat{b}_l$, also breaks $U(1)$ symmetry and describes an asymmetric gain or loss process; i.e. an \textit{incoherent drive}, which could be engineered via coupling to a controlled dissipative environment or measurement setup. 

Importantly, the pairing terms $J_{lm}^x\,\hat{b}_l \hat{b}_m$ and $J_{lm}^x\,\hat{b}^\dagger_l \hat{b}^\dagger_m$ cause many-body eigenstates to become superpositions across different number sectors. This means that in the unperturbed Hermitian basis, the imaginary part of the expectation value $\Im \langle n_{\rm L} | \hat{V}_{\rm NH} | n_{\rm R} \rangle$ can be nonzero, as $|n_{\rm R}\rangle = \sum_i c_i |q_i\rangle_{\rm R}$ includes states with different charges $q_i$. Even weak NH perturbations can therefore introduce complex corrections to the eigenenergies, which helps explaining the enhanced sensitivity to NH effects when the $\hat{\sigma}^x_l \hat{\sigma}^x_m$ term is finite. In this view, removing the pairing and retaining only hopping terms, $\sum_{l,m} J_{lm}^x \left( \hat{b}^\dagger_l \hat{b}_m + \hat{b}_l \hat{b}^\dagger_m \right)$, eliminates first-order corrections to the spectrum. Alternatively, by looking at the form of the non-Hermitian term, one can restore the reality of the spectrum through a valid similarity transformation. This involves choosing a function \(\vartheta \equiv \vartheta(\gamma)\) such that \(\hat{b}_l \to e^{-\vartheta} \hat{b}_l\), \(\hat{b}^\dagger_l \to e^{\vartheta} \hat{b}^\dagger_l\), which maps the pairing terms to \(e^{2\vartheta} \hat{b}_l^\dagger \hat{b}_m^\dagger\) and \(e^{-2\vartheta} \hat{b}_l \hat{b}_m\). Applying the same transformation to the pairing terms restores quasi-Hermiticity and the reality of the spectrum up to a finite critical point $\gamma^c$. In addition, symmetries such as time-reversal can also forbid first-order complex corrections in certain models~\cite{PhysRevLett.123.090603}. In related works, asymmetric hopping models are more commonly studied. In contrast to our model, those typically preserve $U(1)$ symmetry and may exhibit the NH \textit{skin effect} under open boundary conditions~\cite{sannia2024skin}.

As a general result, operators of the form \(\hat{V}_{\rm NH} = \sum_l f(\gamma)\, \hat{\sigma}^+_l + f(-\gamma)\, \hat{\sigma}^-_l\) have a real spectrum as long as \(f(\gamma)f(-\gamma) > 0\). The spectrum becomes complex when this product vanishes and changes sign at the exceptional point at $\gamma^c$. A particular example admitting real eigenvalues regardless of the value of $\gamma$ is when \(f(\gamma) = e^\gamma\)~\cite{PhysRevLett.125.260601}, leading to \(\hat{V}_{\rm NH} = \sum_l e^\gamma\, \hat{\sigma}^+_l + e^{-\gamma}\, \hat{\sigma}^-_l\), which can be written as a similarity transformation of a Hermitian operator as \(\hat{V}_{\rm NH} =\sum_l \hat{S}(\gamma)\, \sx_l \, \hat{S}(\gamma)^{-1}\) with \(\hat{S}(\gamma) = \exp\left( \frac{\gamma}{2} \sum_l \sz_l \right)\). This simply corresponds to \(\vartheta = \gamma/2\). More generally, if \(f(\gamma)\) is not exponential, then the appropriate \(\vartheta(\gamma)\) might involve nonlinear dependency on $\gamma$. Furthermore, for the NH perturbations of the above form we can take $\vartheta\sim \left[\ln f(\gamma)-\ln f (-\gamma)\right]$ and \(\hat{V}_{\rm NH} =\sqrt{f(\gamma)f(-\gamma)}\sum_l \hat{S}(\gamma)\, \sx_l \, \hat{S}(\gamma)^{-1}\) with $S=\exp({\vartheta}\sz)$.

% #######################################################################
% #######################################################################
\section{Evolution of operators for a two-level system}
\label{appendixc}
To elucidate the relation between inputs $\theta$ and the observables, here we consider the time evolution of the operator $\hat{\sigma}^{z}(t)$ in the Heisenberg picture under $ \hat{\mathcal{H}}_{\rm res}=h\sx+i\gamma\sy$ and $\hat{\mathcal{H}}_{\rm rot}=\theta\,{\hat{\sigma}^x}$, expressed as
\begin{align}
    \sz(t)=\frac{\hat{\mathcal{U}}^{\dagger}(t)\,\sz\,\hat{\mathcal{U}}(t)}{{\rm Tr}[\hat{\mathcal{U}}(t)\,\hat{\mathcal{U}}^{\dagger}(t)]},
\end{align}
where $\hat{\mathcal{U}}(t):=\exp{(-it\hat{\mathcal{H}}_{\rm res})}\,\exp{({-i\theta t'}\hat{\sigma}^x)}$. From these we define
\begin{align}
    \hat{{\Sigma}}^z(t):= (c\hat{\mathbb{I}}+is\hat{\mathcal{H}}^{\dagger}_{\rm res}) \,\sz\, (c\hat{\mathbb{I}}-is\hat{\mathcal{H}}_{\rm res}),
\end{align}
where $c:=\cos(\omega t)$, $s:= \sin(\omega t)/\omega$, and $\omega:=\sqrt{h^2-\gamma^2}$. Using the standard relations among the Pauli spin operators, one can easily show that
\begin{align}
   \hat{{\Sigma}}^z(t) = A\,\sz + B\,\sy + C\,\sx+ D\,\mathbb{\hat{I}}.
\end{align}
where
% \begin{align}
% A &= c^2 + s^2(\gamma^2 - 1), & B &= -2cs, \\\nonumber
% C &= 2i\gamma\,cs, & D &= -2\gamma\,s^2.
% \end{align}
\begin{align}
A &= c^2 - s^2(\gamma^2 + h^2), & B &= 2hcs, \\\nonumber
C &= 0, & D &= 2h\gamma\,s^2.
\end{align}
Note also $D=0$ when $\gamma=0$ and/or $h=0$, and $B=0$ only when $h=0$. Close to the exceptional point $\gamma \approx h$, one can confirm that $\hat{\Sigma}^z(t) \approx \left(1 - 2h^2 t^2\right)\, \sz + 2ht \, \sy + 2 h^2 t^2 \, \hat{\mathbb{I}}
$, which is dominated by polynomial (non-oscillatory) terms, a signature of critical NH behavior. Incorporating the input-dependent rotation, and defining $\Theta:=\theta t'$, for the unnormalized evolution one finds
\begin{align}
    \sz(t)&= \hat{\mathcal{U}}_{\rm rot}^\dagger(t')\,\hat{{\Sigma}}^z(t)\, \hat{\mathcal{U}}_{\rm rot}(t')\\ \nonumber &=A\Bigl[\cos(2\Theta)\,\nonumber\sz+\sin(2\Theta)\,\sy\Bigr] \\\nonumber
&+B\Bigl[\cos(2\Theta)\,\sy-\sin(2\Theta)\,\sz\Bigr] \\\nonumber
&+ D\,\mathbb{\hat{I}},
\end{align}
which is a nonlinear function of the inputs. In addition, ${\rm Tr}[\hat{\mathcal{U}}(t)\,\hat{\mathcal{U}}^{\dagger}(t)]=2c^2+2s^2(h^2+\gamma^2)$, which does not depend on the inputs. In the vicinity of the transition point the norm grows quadratically with time ${\rm Tr}[\hat{\mathcal{U}}(t)\,\hat{\mathcal{U}}^{\dagger}(t)]\sim (ht)^2$ and increases exponentially for $\gamma>h$. For small $\Theta$, one finds the linear form $\hat{\sigma}^{z}(t) \approx\left[A - 2\Theta\,B\right]\,\sz+\left[B + 2\Theta\,A\right]\,\sy + D\,\mathbb{\hat{I}}.$
One can also easily see that correlation operators $\hat{\sigma}_l^z(t) \hat{\sigma}^z_m(t)$, even in the single particle case, are a nonlinear (quadratic) functions of inputs. 
 % In a many-body system with global encoding of inputs, the input-output mapping even for small $\Theta$ likely becomes nonlinear~\cite{govia2022nonlinear}. Moreover,
% #######################################################################
% #######################################################################

\section{Distinguishability for a two-level system}
\label{appendixd}

Here we derive a simple expression for the distinguishability $\mathcal{D}(t) = \sqrt{1-|\braket{\uparrow(t)}{\downarrow(t)}|^2}
$ between two pure states under $\hat{\mathcal{H}}_{\rm res}=h\hat{\sigma}^x+i\gamma\hat{\sigma}^y$. The time evolution operator is $\hat{\mathcal{U}}(t)=\exp[-it\hat{\mathcal{H}}_{\rm res}]$, and we have
\begin{align} 
\hat{\mathcal{U}}(t) = \begin{pmatrix}
\cos(\phi) & -i(h+\gamma)\sin(\phi)/{\omega} \\
-i(h-\gamma)\sin(\phi)/\omega & \cos(\phi)
\end{pmatrix},
\end{align}
where we have defined $\phi:=\omega t$ and $\omega:=\sqrt{h^2-\gamma^2}$. From this we can calculate:
\begin{align}
\ket{\uparrow(t)} = \frac{\ket{\psi_\uparrow(t)}}{\sqrt{N_\uparrow(t)}} = \frac{1}{\sqrt{N_\uparrow(t)}} \begin{pmatrix} \cos(\phi) \\ -i(h-\gamma)\sin(\phi)/\omega \end{pmatrix}, \nonumber\\
\ket{\downarrow(t)} = \frac{\ket{\psi_\downarrow(t)}}{\sqrt{N_\downarrow(t)}} = \frac{1}{\sqrt{N_\downarrow(t)}} \begin{pmatrix} -i(h+\gamma)\sin(\phi)/\omega \\ \cos(\phi) \end{pmatrix}.
\end{align}
The (unnormalized) inner product then becomes \(\braket{\psi_\uparrow(t)}{\psi_\downarrow(t)} ={-i\gamma \sin(2\phi)}/{\omega}\). A simple calculation shows that $N_\uparrow(t)N_\downarrow(t)=1+(\gamma^2\sin^2(2\phi)/\omega^2)$. From these it follows that 
\begin{align}
    \mathcal{D}(t)=\left[1+\frac{\gamma^2\,\sin^2\!\Bigl(2\sqrt{h^2-\gamma^2}\,t\Bigr)}{h^2-\gamma^2}\right]^{-1/2},
\end{align}
which is similar to that of Ref.~\cite{PhysRevLett.119.190401}. When $|\gamma|<h$, $\mathcal{D}(t)$ exhibits oscillatory behavior with a period $\propto \omega^{-1} $. For $\gamma\to \infty$ we get $\mathcal{D}(t)\to \left[1+\sinh^2{(2\gamma t)}\right]^{-1/2}\to0$. Near the exceptional point $\gamma\approx h$, the distance asymptotically behaves as $\mathcal{D}(t)\sim t^{-1}$. The exponent of this power-law decay depends both on the model and the initial state considered~\cite{PhysRevLett.123.230401}.
% #######################################################################
% #######################################################################

\section{Generating non-Hermitian dynamics\label{sec:generating_non_hermitian}}

As we have demonstrated, non-Hermiticity endows a quantum reservoir with highly desirable properties for information processing, including controllable dissipation and purification. A key question is how to \emph{realize} such NH behavior in a \emph{controlled} manner on a quantum device.

A central insight is that any effective NH Hamiltonian can be generated by continuously monitoring a system, followed by \emph{postselection} on a particular set of measurement outcomes~\cite{ashida2020non,PhysRevB.108.134305}. Concretely, suppose a system evolves unitarily under a Hermitian Hamiltonian $\hat{\mathcal{H}}$ but is also subjected to repeated weak measurements described by local jump operators $\{\hat{\mathcal{L}}_l\}$. Postselecting the ``no-click'' trajectories (i.e., the measurement record with no quantum jumps) removes all events that would have led to a projection via $\hat{\mathcal{L}}_l$, leaving behind a deterministic, nonunitary evolution effectively generated by
\begin{equation}
\hat{\mathcal{H}}_{\rm res} \;=\;\hat{\mathcal{H}} \;-\; \frac{i\gamma}{2}\sum_l \hat{\mathcal{L}}_l^\dagger \hat{\mathcal{L}_l}.
\label{eq:H_res_eff}
\end{equation}
While this perspective captures the physics of measurement-induced non-Hermiticity, it does not necessarily simplify the \emph{implementation} of such dynamics in a programmable quantum device, where one often seeks direct engineering of gain/loss.

A constructive approach for emulating NH evolutions employs the fact that any infinitesimal evolution over a small time step $\delta t$ can be decomposed into a (normalised) sum of unitaries. These unitary evolutions can then be sampled (or deterministically combined) to reproduce the desired nonunitary effect~\cite{PhysRevResearch.3.013017,QDE}. One version of this idea, called \emph{ensemble rank truncation} (ERT)~\cite{PhysRevResearch.3.013017}, decomposes a Lindbladian evolution into an ensemble of pure-state trajectories, then applies a principal-component truncation to keep the representation efficient. More recently, \emph{quantum dynamical emulation} (QDE)~\cite{QDE} has been proposed as a technique for mapping nonunitary operators (including imaginary-time evolutions and more general dissipative processes) onto a carefully chosen set of unitaries. Here, one evolves forward under each of these unitaries, combines the results in a weighted fashion (which can be done purely as a classical post-processing step on measurement outcomes), and thereby reproduces the effect of a NH Hamiltonian via a prescribed ensemble average of unitaries.

We now illustrate how such a decomposition arises when trying to emulate the effective Hamiltonian of Eq.~\eqref{eq:H_res_eff}. Presuming at time $\hat{\rho}(t)$ is normalised, for an infinitesimal time step $\delta t$, the unnormalized density matrix $\bar{\rho}(t+\delta t)$ evolves as
\begin{align}
  \bar{\rho}(t+\delta t)
  \;=&\;\hat{\rho}(t)\;-\; i\,\delta t\,\bigl[\hat{\mathcal{H}},\hat{\rho}(t)\bigr] \notag \\
  \;-&\;\delta t\,\frac{\gamma}{2}\sum_l\!\bigl\{\hat{\rho}(t),\hat{\mathcal{L}}_l^\dagger \hat{\mathcal{L}}_l\bigr\}
  \;+\;\mathcal{O}(\delta t^2),
\label{eq:infinitesimal_dynamics}
\end{align}
where $\{\hat{A},\hat{B}\}$ is the anticommutator. Noting that  $\hat{\mathcal{L}}_l^\dagger \hat{\mathcal{L}}_l$ is guaranteed to be Hermitian, we first introduce a shift $\bar{\lambda}_l$, where
\begin{equation}
\bar{\lambda}_l > \min[|\lambda_l|^2]
\end{equation}
and $\{\lambda_l\}$ are the eigenvalues of $\hat{\mathcal{L}}_l$. With this shift, we can assign 
\begin{equation}
\bar{\mathcal{L}}^\dagger_l\bar{\mathcal{L}}_l = \hat{\mathcal{L}}_l^\dagger \hat{\mathcal{L}}_l +\bar{\lambda}_l \hat{1}.
\end{equation}
The effect of this is to render the spectrum of $\bar{\mathcal{L}}^\dagger_l\bar{\mathcal{L}}_l$ positive, such that $\bar{\mathcal{L}}_l$ is a Hermitian operator. Notably, the effect of this shift is a scaling of the normalisation of the infinitesimally evolved density. That is, if we shift $\hat{\mathcal{L}}_l \to \bar{\mathcal{L}}_l$ then $\bar{\rho}(t+\delta t)$ is simply scaled by a factor ${\rm exp}[\gamma\delta t \sum_l \bar{\lambda}_l]$. Consequently we may freely shift the spectrum of the nonunitary part of the dynamical generator, and absorb its effect on the dynamics into the normalisation of the density. 

Equipped with the Hermitian analogue $\bar{\mathcal{L}}_l$, we now define a family of \textit{effective} Hamiltonians $\hat{\mathcal{H}}_{\pm l}$: 
\begin{equation}
  \hat{\mathcal{H}}_{\pm l} \;=\;\hat{\mathcal{H}}\;\pm\;\sqrt{\frac{L\,\gamma}{2\,\delta t}}\;\bar{\mathcal{L}}_l,
\end{equation}
where $L$ is the total number of jump operators. For each of these there is an associated unitary propagator denoted by $\hat{\mathcal{U}}_{\pm l} \;=\; \exp\bigl[-\,i\,\delta t\,\hat{\mathcal{H}}_{\pm l}\bigr]$. With these, Eq.\eqref{eq:infinitesimal_dynamics} can be approximated to order $\delta t^2$ via
\begin{equation}
    \bar{\rho}(t+\delta t)
    \;\approx\;
    \frac{1}{2\,L}\;
    \sum_{l}\!
    \Bigl(\hat{\mathcal{U}}_{+ l} + \hat{\mathcal{U}}_{-l}\Bigr)\,
    \hat{\rho}(t)\,
    \Bigl(\hat{\mathcal{U}}_{+ l} + \hat{\mathcal{U}}_{-l}\Bigr)^\dagger.
\end{equation}
 Tracing over both sides gives the normalisation of this ensemble evolution:
\begin{equation}
    {\rm{Tr}}[\bar{\rho}(t+\delta t)]=1+\frac{1}{2L}\sum_l {\rm Tr}\left[\hat{\rho}(t) \hat{\mathcal{U}}^\dagger_{+ l}\hat{\mathcal{U}}_{-l} + {\rm h.c}.\right] 
\end{equation}
Evaluating the RHS leads to (up to order $\delta t^2$)
\begin{equation}
{\rm{Tr}}[\bar{\rho}(t+\delta t)]={\rm e}^{-\gamma \delta t\Gamma}
\end{equation}
where $\Gamma=\sum_l {\rm Tr}[\hat{\rho}(t) \bar{\mathcal{L}}_l^\dagger\bar{\mathcal{L}}_l]$. This normalization factor is consistent with the predicted decay implied by Eq. \eqref{eq:ures_map}, together with the additional scaling required to complement the $\bar{\lambda}_l$ shift. Restoring normalization then provides a short-time update rule:
\begin{equation}
    \hat{\rho}(t+\delta t)
    \;\approx\;
    \frac{
    {\rm e}^{\gamma\,\delta t\,\Gamma}
    }{2\,L}\;
    \sum_{l}\!
    \Bigl(\hat{\mathcal{U}}_{+l} + \hat{\mathcal{U}}_{-l}\Bigr)\,
    \hat{\rho}(t)\,
    \Bigl(\hat{\mathcal{U}}^\dagger_{+l} + \hat{\mathcal{U}}^\dagger_{-l}\Bigr).
\label{eq:rho_update_approx}
\end{equation}
Hence, \emph{any} desired NH evolution of the form~\eqref{eq:H_res_eff} can be modeled at each step by an ensemble of purely \emph{unitary} updates. The advantage of this construction is that NH dynamics can  be \emph{emulated} constructively by a suitable collection of unitaries, rather than by explicitly implementing gain/loss or discarding measurement records. Moreover, the deterministic or ensemble-based viewpoint allows recently developed techniques such as Ensemble Rank Truncation (ERT)~\cite{PhysRevResearch.3.013017} and Quantum Dynamical Emulation (QDE)~\cite{QDE} to be applied, greatly facilitating the controlled realization of NH physics. Notably, QDE has already been experimentally demonstrated for an interacting spin system’s imaginary-time evolution in a superconducting quantum processor. Consequently the advantages conferred by NH dynamics for quantum information processing can be obtained without requiring a \textit{physical} implementation of non-Hermiticity.

\bibliography{ref.bib}
\bibliographystyle{apsrev4-1}
\end{document}